\documentclass[journal]{IEEEtran}
\IEEEoverridecommandlockouts
\usepackage{cite}
\usepackage{amsmath,amssymb,amsfonts}
\usepackage{algorithmic}
\usepackage{array}
\usepackage{graphicx}
\usepackage{textcomp}
\usepackage{stfloats}
\usepackage{url}
\usepackage{verbatim}
\usepackage{graphicx}
\usepackage{xcolor}
\usepackage[switch]{lineno} 
\def\BibTeX{{\rm B\kern-.05em{\sc i\kern-.025em b}\kern-.08em
    T\kern-.1667em\lower.7ex\hbox{E}\kern-.125emX}}
\begin{document}
\title{Adaptive LPD Radar Waveform Design\\ with Generative Deep Learning\\}  

\author{Matthew R. Ziemann and Christopher A. Metzler
\thanks{Manuscript received July 9, 2024; revised January 2, 2025. Part of this work was presented at the 2023 IEEE Asilomar Conference on Signals, Systems, and Computers \cite{Ziemann2023}. \textit{(Corresponding Author: Matthew Ziemann)}}
\thanks{M. R. Ziemann is with the DEVCOM Army Research Lab, Adelphi, MD, USA, and with the Department of Computer Science, University of Maryland, College Park, MD, USA. E-mail: matthew.r.ziemann.civ@army.mil.}
\thanks{C. A. Metzler is with the Department of Computer Science, University of Maryland, College Park, MD, USA. E-mail: metzler@umd.edu.}
\thanks{Digital Object Identifier 10.1109/TRS.2025.3542283}}

\maketitle

\begin{abstract}
We propose a learning-based method for adaptively generating  low probability of detection (LPD) radar waveforms that blend into their operating environment. 
Our waveforms are designed to follow a distribution that is indistinguishable from the ambient radio frequency (RF) background---while still being effective at ranging and sensing. To do so, we use an unsupervised, adversarial learning framework; our generator network produces waveforms designed to confuse a critic network, which is optimized to differentiate generated waveforms from the background. 
To ensure our generated waveforms are still effective for sensing, we introduce and minimize an ambiguity function-based loss on the generated waveforms.
We evaluate the performance of our method by comparing the single-pulse detectability of our generated waveforms with traditional LPD waveforms using a separately trained detection neural network. We find that our method can generate LPD waveforms that reduce detectability by up to 90\% while simultaneously offering improved ambiguity function (sensing) characteristics. 
Our framework also provides a mechanism to trade-off detectability and sensing performance.
\end{abstract}


\section{Introduction}
\label{section::Intro}

\IEEEPARstart{R}{adar} systems are critical for military operations. They allow for surveillance, reconnaissance, navigation, fire control, air defense, and a host of other tasks on platforms that range from small unmanned systems to massive aircraft carriers. However, radar systems are vulnerable to detection and counter-action by adversary electronic support systems, including radar warning receivers and anti-radiation missiles \cite{IntroEW}. Low probability of detection (LPD) radar addresses this vulnerability through the use of specialized waveforms and emission patterns that lower their likelihood of detection \cite{Liu2001}. 

LPD waveforms are designed to minimize detectability while maintaining necessary radar performance metrics such as range and velocity resolution. Traditional methods for LPD radar waveform design include low peak power waveforms, wideband waveforms, frequency agile or hopping waveforms, coded or modulated waveforms, and irregular scan patterns \cite{RFStealth}. These methods aim to reduce the peak power, bandwidth, or duration of the radar signal---or to introduce randomness or complexity in the signal structure---so intercept receivers cannot easily detect or identify the radar signal. However, these waveforms may still be vulnerable to advanced intercept systems that can exploit their statistical features, such as cyclostationarity \cite{Lime2002, Liu2022}. Modern software defined radar (SDRadar) has opened the door to increased waveform diversity---especially for cognitive radar applications---but the development of algorithms for adaptive SDRadar systems is still in its infancy \cite{Blunt2016, Martone2021a, Martone2021}.

In recent years, deep learning has enabled significant advances in radar signal processing, including advanced modulation classification, waveform design, and synthetic data generation \cite{RFML1,Pan2023,RFML2,RFML3,Lee2023}. We propose a novel application of deep learning for LPD radar waveform design. Generative machine learning techniques, such as generative adversarial networks (GANs), can learn the underlying distribution of a dataset and generate new examples that closely match the original distribution \cite{GAN, Ruthotto2021}. This makes them well-suited for LPD waveform design tasks, as they can learn the statistical characteristics of the environment and generate radar waveforms that mimic background radio frequency (RF) signals.

\begin{figure}
  \centering
  \includegraphics[width=\columnwidth]{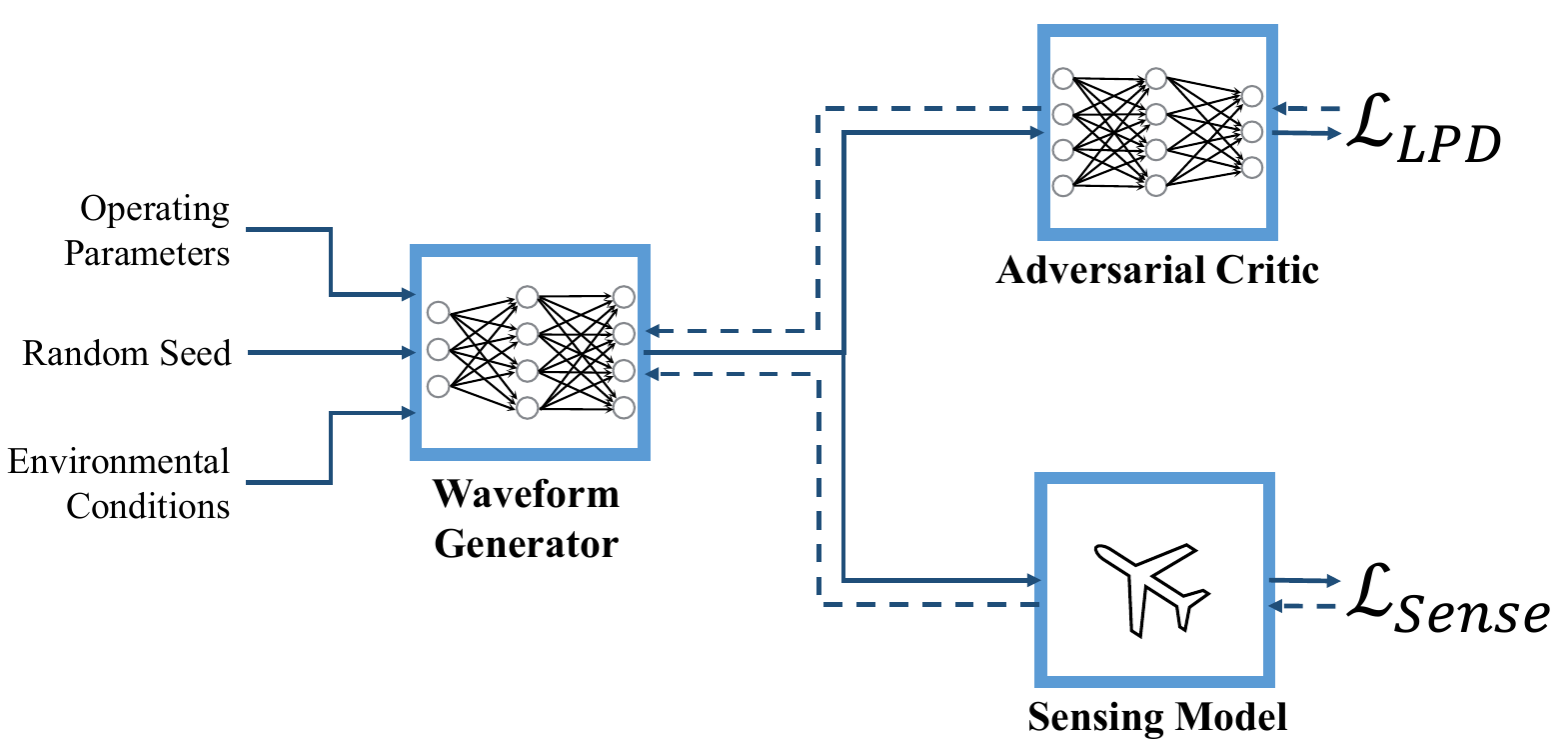}
  \caption{\textbf{LPD Radar Waveform Generation Framework.} 
  The generative model optimizes range/velocity resolution while minimizing detectability.}
  \label{overview}
  \vspace{-5pt}
\end{figure}

In this work, we employ unsupervised GANs to generate unique and adaptive radar waveforms that match the statistical distribution of the background RF environment. In particular, we train a conditional Wasserstein GAN to generate LPD radar waveforms based on the distribution of common RF communication modulations. We utilize an objective based on a differentiable ambiguity function to optimize its outputs for range and Doppler ambiguity requirements. An overview of this framework is shown in Fig. \ref{overview}. We find that our generated waveforms exhibit up to a 90\% reduction in detectability compared to other LPD waveforms while maintaining useful range and velocity resolutions. In addition, we can increase the strength of the ambiguity function-based objective to further improve ambiguity characteristics at the cost of increased detectability.

\section{Related Work}
\label{section::PriorWork}
Radar waveforms determine how well a radar system can detect, locate, and classify objects; how much interference a radar system causes and how sensitive a system is to interference and jamming; and how easy it is for a third party to detect a radar system. The primary objective of radar waveform design is to optimize these trade-offs so as to ensure a system can sense, coexist with, and survive in its environment.\\

\noindent \textbf{Waveform Design.}
Over the years, the field of radar waveform design has evolved significantly, driven by both technological advancements and changing operational requirements. Initial designs were largely constrained by hardware limitations, focusing on frequency-modulated and phase-coded pulses \cite{Levanon2004}. As technology progressed, more complex designs emerged, such as pulse-Doppler, frequency-modulated continuous-wave (FMCW), and multiple-input, multiple-output (MIMO) systems \cite{Melvin2013}. These advancements helped to expand the operational scope of radar systems, allowing for more accurate and reliable surveillance in a variety of conditions. However, these waveform designs are primarily based on static or predefined scenarios and have limited adaptability in dynamic environments.\\

\noindent \textbf{Low Probability of Detection.} 
Waveform design for Low Probability of Detection (LPD) and Low Probability of Intercept (LPI) has been an area of significant interest, particularly for military applications. The primary goal of these designs is to minimize the radar's detectability while maintaining sufficient operational performance. Early LPI/LPD waveforms were characterized by low temporal, spectral, or spatial energy densities, which made them difficult to discern from background noise \cite{Stove2004}. Techniques such as FMCW, frequency hopping, and non-traditional modulation methods like polyphase coding were common \cite{RFStealth}. However, these methods often came with trade-offs in terms of effective range, resolution, or other factors such as complexity and cost \cite{RFStealth}.
Noise radar emerged as an additional LPD approach that utilized random noise-like signals to reduce detectability. These waveforms have desirable ambiguity functions and are resistant to electronic countermeasures (i.e., jamming), but their random nature requires longer dwell times and more expensive processing \cite{Narayanan2016}.\\

\noindent \textbf{Cognitive Radar.} 
The concept of cognitive radar emerged in parallel with advancements in waveform design as a means of rapidly adapting radar to changing conditions \cite{Gurbuz2019}. Early cognitive radar, defined by the \textit{sensor management} paradigm, used decision algorithms to adjust radar emission parameters such as power, direction, and beamwidth \cite{Hero2011}. This quickly evolved to incorporate adaptive signal processing (e.g., space-time adaptive processing) \cite{Ward1998}---and, eventually, waveform selection and intra-pulse adaptation \cite{Kershaw1994, Kim2010}. This growing field of \textit{waveform diversity} \cite{Blunt2016} aims to optimize the radar waveform itself to maximize performance according to particular scenarios and tasks \cite{IEEE686}. 

Modern waveform diversity incorporates the perception-action cycle (PAC) to optimize emissions based on the radar's environment \cite{Haykin2012}. In this framework, the cognitive system is provided with measurements from its environment in order to intelligently select emission parameters (e.g., waveform class, pulse repetition interval, and bandwidth \cite{Kim2010}). While a host of techniques have been explored to accomplish this in applications varying from target detection to spectrum sharing \cite{NATO227}, there are still many open research questions regarding optimization objective, technique selection, algorithm response time, and more \cite{Martone2021a, Martone2021}. For example, the extension of these techniques to LPD objectives---such as adaptive spectrum shaping of noise radar \cite{Pici2019}---and the incorporation of emerging deep learning and reinforcement learning methodologies \cite{Thornton2020} are two emerging areas of interest.\\

In this work, we introduce a novel application of unsupervised generative adversarial networks (GANs) to adaptively generate optimal LPD radar waveforms from measurements of the RF background. By using a conditional GAN, we simplify the perception-action cycle by generating waveforms directly from measurements. Unlike supervised waveform design methods, we do not train our GAN on a dataset of radar waveforms \cite{Saarinen2023, RFML3}. Instead, we train our GAN on unlabeled data that is representative of (or taken directly from) the target environment, and then fine-tune it with a user-defined ambiguity function to produce LPD waveforms with desirable ambiguity characteristics. These LPD waveforms match the statistical behavior of the RF background distribution, rather than relying on assumptions about the environment or detector to reduce detectability (e.g., white noise, frequency hopping). This flexible framework reduces the data burden associated with deep learning-based techniques while enabling the model to adapt to changing RF environments or operational requirements.\\


\section{Problem Formulation}
\label{section::Problem}
We assume a signal model of the form
\begin{align}
    x(t) &= \alpha s(t) + n(t) + b(t), \\
    \alpha &= \sqrt{\frac{G_t G_i \lambda^2 L}{(4 \pi )^2 R^2}},
\end{align}
where $x(t)$ is the detector's received signal, $s(t)$ is the emitted radar signal, and $\alpha$ is the energy attenuation coefficient defined by the transmitter and detector gain $G_{t,i}$, wavelength $\lambda$, path loss $L$, and distance between the the emitter and detector $R$. Noise is denoted with $n(t)$, and $b(t)$ denotes the RF background signal(s) present in the spectrum. We assume the background $b(t)$ follows a distribution $\mathbb{P}_b$ and that we can observe many realizations  $b_1(t), b_2(t), ..., b_m(t)$ of background.

\subsection{Waveform Design} 
\label{section::WaveformDesign}

\begin{figure}[t]
    \centerline{\includegraphics[width=0.9\columnwidth]{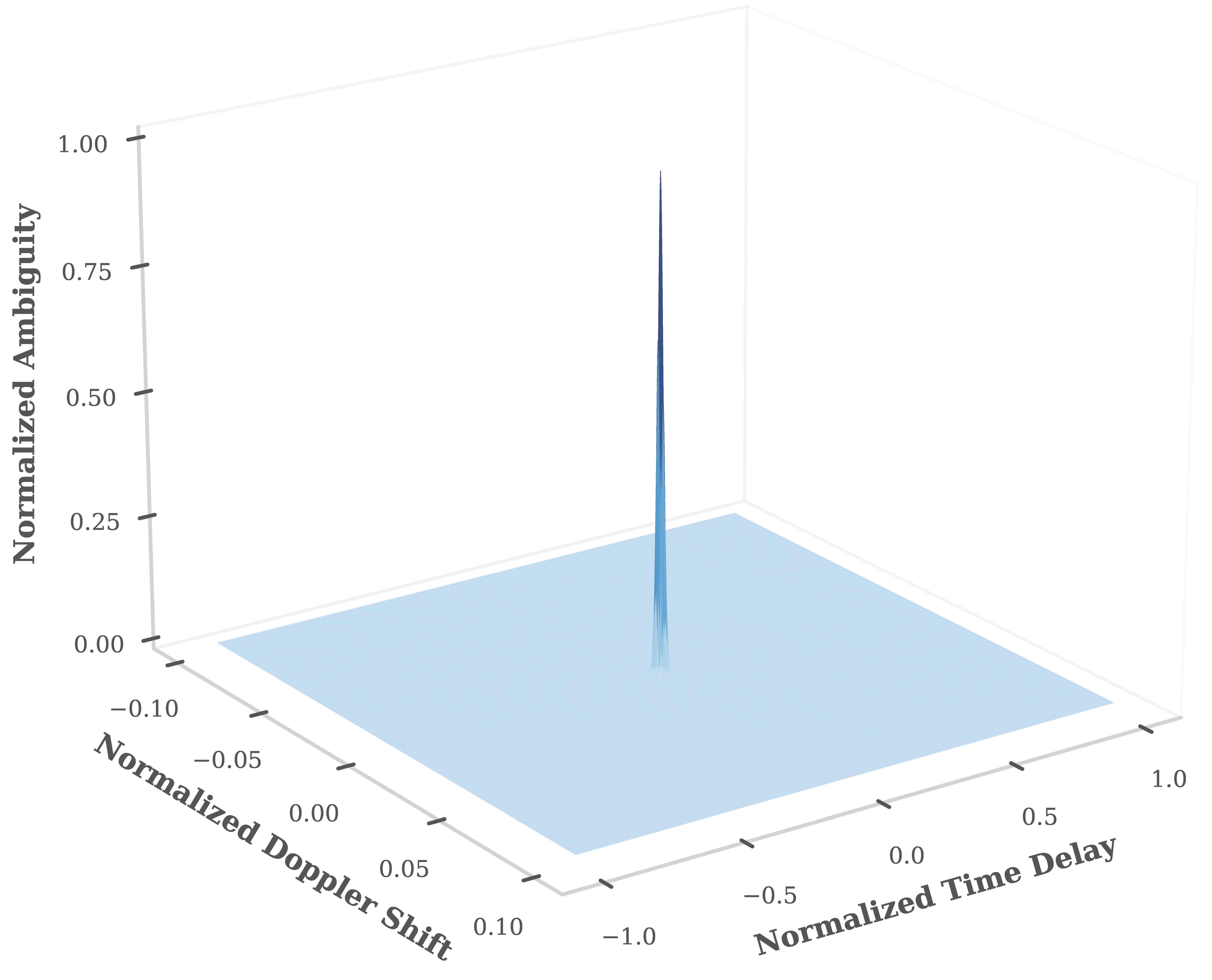}}
    \caption{\textbf{Thumbtack Ambiguity Function.} Example of an optimal ``thumbtack'' ambiguity function with low delay (range) and Doppler (velocity) ambiguities.}
    \label{thumbtack}
\end{figure}

Given this problem setting, our goal is to find a signal $s(t)$ that has good radar performance and is difficult to detect. Radar performance is well described by the \textit{ambiguity function}, which computes the matched filter response for a signal under both time and Doppler mismatches \cite{Richards2022}. The complex ambiguity function $\hat{A}$ for time shift $\tau$ and Doppler shift $F_D$ is
\begin{equation}
\hat{A}(\tau, F_D) = \int_{-\infty}^{\infty} s(t)\exp(j2\pi F_D t) s^{\ast}(t-\tau)dt.
\end{equation}
This can be written in terms of the signal spectrum as
\begin{equation}
\hat{A}(\tau, F_D) = \int_{-\infty}^{\infty} S^{\ast}(F)S(F-F_D)\exp(j2\pi F \tau)dF,
\end{equation}
where $S(F)$ represents the Fourier transform of the time domain signal $s(t)$. The ideal ambiguity function is dependent on the radar task, but desirable characteristics often include a narrow mainlobe and low ambiguity in both delay (range) and Doppler (velocity). This is often described as the ``thumbtack'' ambiguity function depicted in Fig. \ref{thumbtack}.

Optimizing a waveform for low detectability is more difficult, as there is not a singular definition of detectability. Peak-to-average power ratio, instantaneous bandwidth, and variable transmit period are just a few of many variables affecting detectability \cite{RFStealth}. Chen \textit{et al.} \cite{Chen2022-a} introduce a novel approach for assessing LPD by utilizing the Kullback--Leibler divergence (KLD) between a radar waveform and the RF background. The KLD quantifies the difference between two distributions, and reducing the KLD between the radar waveform and RF background increases their similarity and reduces detectability. For some measurable space $\chi$ with probability measures $\text{Prob}(\chi)$, we can define the KLD as
\begin{equation}
KL(\mathbb{P}_b\|\mathbb{P}_g) = \int \log \biggl( \frac{P_b(x)}{P_g(x)} \biggr) P_b(x)d\mu (x),
\end{equation}
where $\mathbb{P}_b$ and $\mathbb{P}_g$ are the distributions of RF background and generated signals such that $\mathbb{P}_b, \mathbb{P}_g \in \text{Prob}(\chi)$, $\mu$ is some measure defined on $\chi$, and $P_b$ and $P_g$ are densities with respect to $\mu$.

A good LPD waveform is therefore one that is similar to the RF background distribution and has a desirable ambiguity function. We can set up this optimization problem by minimizing the distance between the distributions of generated and background signals, and by minimizing the distance between the generated ambiguity functions and some desired target ambiguity function. In this work, we accomplish this optimization with deep neural networks.

\begin{figure*}[t]
    \centering
    \begin{minipage}{0.49\textwidth}
        \centering
        \footnotesize{\textbf{Generator}}\\
        \includegraphics[width=\textwidth]{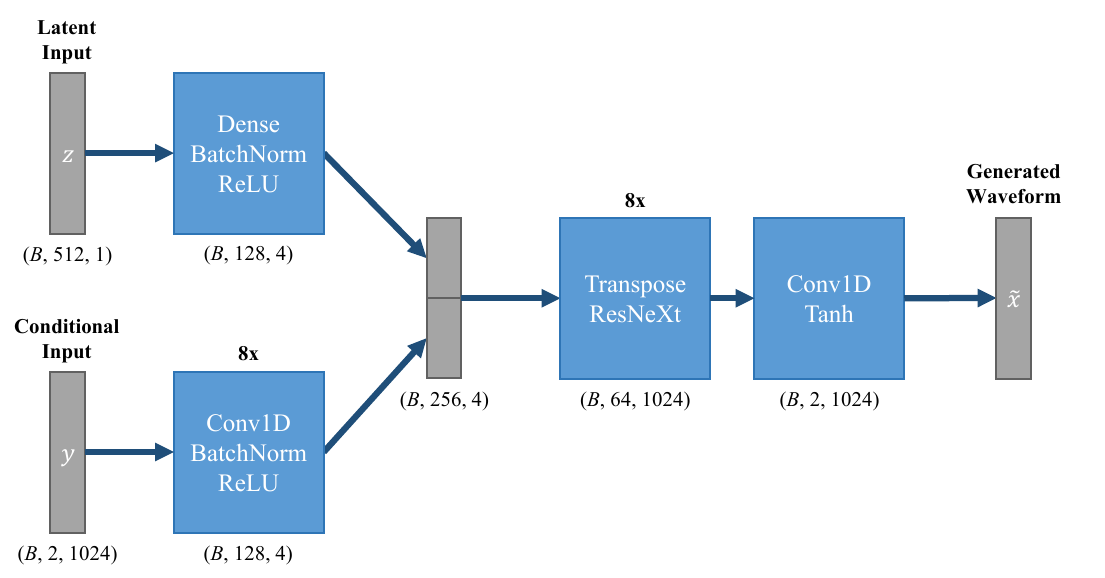}
        \label{fig:generator}
    \end{minipage}
    \hfill  
    \begin{minipage}{0.49\textwidth}
        \centering
        \footnotesize{\textbf{Critic}}\\
        \includegraphics[width=\textwidth]{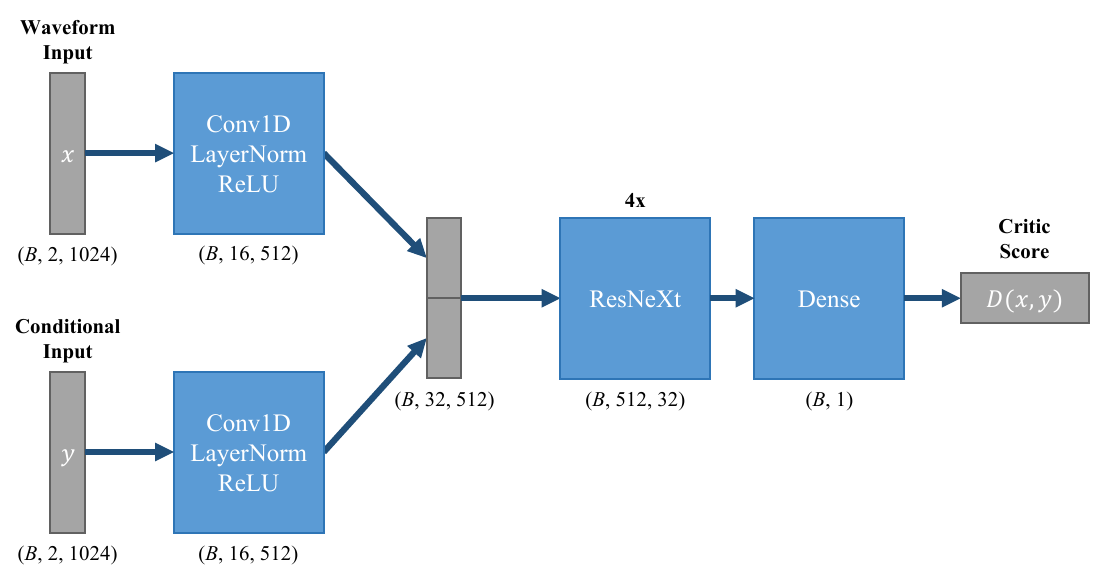}
        \label{fig:critic}
    \end{minipage}
    \caption{\textbf{Network Architectures.} Overview diagrams of the generator network (left) and critic network (right) with 1024 signal length and 512 latent vector length. Output shapes are shown below layers with batch size $B$.}
    \label{fig:architectures}
\end{figure*}

\subsection{Generative Adversarial Networks} 
\label{section::Generative}
Deep generative models (DGMs) are a type of machine learning model that aim to learn the underlying probability distribution of complex and high-dimensional data \cite{Ruthotto2021}. They can then generate new and representative samples from the learned distribution. For example, a DGM trained on an image dataset of human faces can generate realistic faces that do not actually exist \cite{DiffusionModels}. Common DGMs include variational autoencoders (VAEs) \cite{VAE}, generative adversarial networks (GANs) \cite{GAN}, normalizing flows \cite{RealNVP}, and diffusion models \cite{DiffusionModels}. Although each method has strengths, we utilize GANs in this work as the adversarial training framework is well suited to our LPD waveform design problem---the framework is explicitly training a generator to trick a detector, which closely mirrors the LPD challenge.

GANs are composed of two separate neural networks---a generator and a discriminator---trained in a zero-sum game. The generator produces fake examples from a random noise vector, and the discriminator distinguishes between fake examples and real examples from the training data distribution (sometimes called true or ``target'' examples). The networks are pitted against one another, and, as they train, the generator learns to produce examples that are indistinguishable from the real ones. Together, these networks optimize the following minimax objective:
\begin{equation}
    \label{eq:GAN_Objective}
    \min_G \max_D \mathop{\mathbb{E}}_{x \sim \mathbb{P}_r}[\log D(x)] 
    + \mathop{\mathbb{E}}_{\tilde{x} \sim \mathbb{P}_g}[\log (1-D(\tilde{x}))],
\end{equation}
where $G$ is the generator, $D$ is the discriminator, $\mathbb{P}_r$ is the real data distribution, and $\mathbb{P}_g$ is the generated data distribution defined by generated data $\tilde{x} = G(z)$. Here, $z$ is the input to the generator that is randomly sampled from some distribution (e.g., Gaussian). 

As these networks train, they effectively minimize the Jensen--Shannon divergence (JSD) between the real and generated distributions \cite{Arjovsky2017}:
\begin{equation}
JS(\mathbb{P}_r \| \mathbb{P}_g) = \frac{1}{2} KL(\mathbb{P}_r \| \mathbb{P}_A) + \frac{1}{2} KL(\mathbb{P}_g \| \mathbb{P}_A),
\end{equation}
where $\mathbb{P}_A$ is the average distribution $\frac{P_r + P_g}{2}$. JSD is based on KLD, though it is generally easier to optimize because it is symmetric and always finite. However, this objective is not always continuous and leads to inconsistent training. The Wasserstein distance is a more suitable alternative that is continuous everywhere, and so we use the Wasserstein GAN (WGAN) \cite{WGAN}. Rather than minimizing the JSD, WGAN minimizes the Wasserstein-1 distance (also called the Earth-Mover distance):
\begin{equation}
W(\mathbb{P}_r \| \mathbb{P}_g) = \inf_{\gamma \in \Pi(\mathbb{P}_r, \mathbb{P}_g)} 
\mathbb{E}_{(x,y) \sim \gamma} {[ \|x-y\|]},
\end{equation}
where $\Pi(\mathbb{P}_r, \mathbb{P}_g)$ is the set of all joint distributions $\gamma(x,y)$ with marginals $\mathbb{P}_r$ and $\mathbb{P}_g$. Using the Kantorovich--Rubinstein duality, the GAN objective \eqref{eq:GAN_Objective} can then be rewritten to:
\begin{equation}
    \label{eq:WGAN_Objective}
    \min_G \max_{D\in \mathcal{D}} \mathop{\mathbb{E}}_{x \sim \mathbb{P}_r}[D(x)] 
    - \mathop{\mathbb{E}}_{\tilde{x} \sim \mathbb{P}_g}[D(\tilde{x})],
\end{equation}
where $\mathcal{D}$ is the set of 1-Lipschitz functions. In WGANs, the discriminator is instead referred to as a critic. It no longer outputs a probability of an example being ``real'' or ``fake,'' but provides a continuous valued score used to compute the Wasserstein distance between distributions. This improves training stability and allows the critic to be a more meaningful measure of output quality. The 1-Lipschitz constraint is enforced with a gradient penalty term to constrain the norm of the network's gradients to at most 1 for a random subset of examples $ \hat{x} \sim \mathbb{P}_{\hat{x}} $ \cite{WGAN-GP}. This new objective for WGAN with gradient penalty (WGAN-GP) is
\begin{equation}
    \begin{aligned}
    \min_{G} \max_{D} &\mathop{\mathbb{E}}_{x \sim \mathbb{P}_r}[D(x)] 
    - \mathop{\mathbb{E}}_{\tilde{x} \sim \mathbb{P}_g}[D(\tilde{x})] \\
    & + \lambda \mathop{\mathbb{E}}_{\hat{x} \sim \mathbb{P}_{\hat{x}}}[(\| \nabla_{\hat{x}} D(\hat{x})\|_2 -1)^2]. 
    \end{aligned} 
\label{eq:WGAN-GP}
\end{equation}

By optimizing \eqref{eq:WGAN-GP}, we force the generator's distribution $\mathbb{P}_g$ to match the real distribution $\mathbb{P}_r$. We can use this Wasserstein-1 minimization in place of KLD for LPD waveform generation by setting the ``real'' distribution equal to the empirical RF background distribution ($\mathbb{P}_r = \mathbb{P}_b$) for a target environment. This will force our generator to learn to generate waveforms that are statistically similar---have small Wasserstein-1 distance---to the RF background.

In the context of RF background signals, it is crucial to note that the statistical properties of the distribution \(\mathbb{P}_b\) are time-dependent. Take, for example, a multi-emitter urban environment.
Over long timescales---i.e., minutes to hours---the background exhibits consistent, cyclical behaviors \cite{Breton2019}. However, over shorter timescales---i.e., microseconds to milliseconds---the background is inherently nonstationary, with rapidly fluctuating characteristics as overlapping, independent signals are emitted by various sources \cite{Ricciardi2021}. To reconcile these disparate timescales and to adapt to the nonstationary nature of short-term RF background signals, we employ a conditional WGAN-GP (cWGAN-GP) model. By doing so, we enable the generator to produce waveforms from the conditional distribution \(\mathbb{P}_{g|y}\), where \(y\) is an instantaneous RF background measurement taken at or near the time of inference. This conditional approach allows us to train the WGAN on a long-term, quasi-stationary distribution while generating examples that are instantaneously conditioned on the most recent sample, thereby providing a more accurate Wasserstein-1 approximation in the nonstationary regime. This gives us the cWGAN-GP objective:
\begin{equation}
    \begin{aligned}
    L_W = \min_{G} \max_{D} &\mathop{\mathbb{E}}_{x \sim \mathbb{P}_b}[D(x|y)] 
    - \mathop{\mathbb{E}}_{\tilde{x} \sim \mathbb{P}_{g|y}}[D(\tilde{x}|y)] \\
    & + \lambda \mathop{\mathbb{E}}_{\hat{x} \sim \mathbb{P}_{\hat{x}}}[(\| \nabla_{\hat{x}} D(\hat{x}|y)\|_2 -1)^2], 
    \end{aligned} 
\label{eq:cWGAN-GP}
\end{equation}

\noindent where $\mathbb{P}_{g|y}$ is the conditional generated data distribution defined by $\tilde{x} = G(z|y)$. In terms of our generator and critic networks, this conditional term is simply an additional input to the networks. This enables the generator to have information of the instantaneous RF background at inference time to improve its output.

\section{Methodology} 
\label{section::Methodology}

We formulate the LPD waveform design problem as jointly maximizing similarity to the RF background and similarity to a desired ambiguity function. We accomplish this with two distinct objectives in our WGAN. First, we use the cWGAN-GP loss function to force the distribution of generated waveforms to be similar to a defined RF background. Second, we use an ambiguity loss term to optimize the ambiguity of generated waveforms toward a target ambiguity function. The resulting generator neural network is then capable of producing a diverse set of LPD radar waveforms, conditioned on the instantaneous RF background, with desirable ambiguity characteristics.

\subsection{Wasserstein GAN} 
\label{section::WGAN}

 We optimize our LPD and ambiguity objectives with a conditional Wasserstein GAN with gradient penalty (cWGAN-GP) implemented in PyTorch \cite{pytorch}. The generator and critic in our cWGAN-GP are based on a 1D variant of the ResNeXt architecture \cite{ResNeXt}. ResNeXt is a modular architecture built from \textit{residual blocks} that utilize parallel convolution pathways and a residual (``skip'') connection. The \textit{cardinality} of the residual layer defines the number of parallel pathways. Importantly, each pathway has a lower dimensionality than a standard ResNet layer. This enables a greater degree of expressivity without increasing computational complexity. Each convolution is followed by batch normalization and a rectified linear unit (ReLU) activation function. We replace batch normalization with layer normalization in the critic to prevent correlation between examples, which is incompatible with our gradient penalty \cite{WGAN-GP}.

An overview of our network architectures can be seen in Fig. \ref{fig:architectures}. We utilize IQ data with real and imaginary components separated into two channels. The generator outputs a 1024-sample IQ output from its latent ($z$) and conditional ($y$) inputs, and it uses a $2\tanh (x/2)$ activation function to ensure its outputs are close to the range of the training data. The critic outputs a critic score from its waveform and conditional ($y$) inputs, where the waveform can be either a real ($x$) or generated signal ($\tilde{x}$). We select conditioning waveforms ($y$) that are the same class and SNR as their $x$ counterparts. In a more realistic scenario, $y$ could be chosen to be any examples that are temporally close to $x$ (e.g., two signals captured in quick succession). The generator and critic have 12.2 and 11.7 million trainable parameters, respectively.

\begin{figure}[t]
    \centerline{\includegraphics[width=\columnwidth]{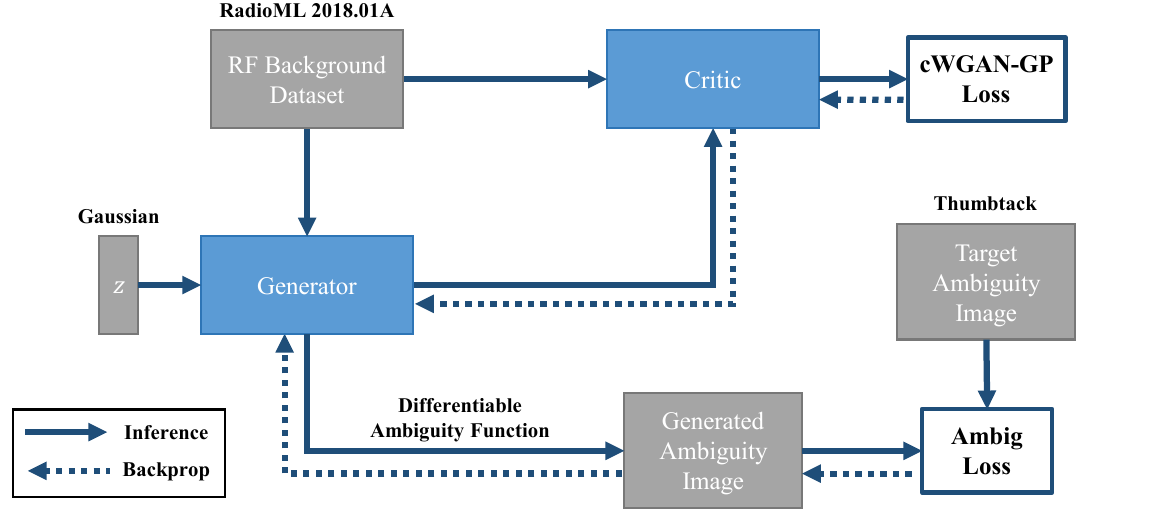}}
    \caption{\textbf{Adversarial Training Framework.} Training process for our cWGAN-GP, where our generator learns to produce waveforms with desirable ambiguity functions that mimic the RF background. Training is initially performed without the ambiguity loss, and alternates between training the critic and the generator. Once converged, the generator is fine-tuned for a small number of epochs with both the cWGAN-GP loss and ambiguity loss.}
    \label{Training_Framework}
\end{figure}

 We train the generator and critic with the method shown in Fig. \ref{Training_Framework}. We alternate training the generator and critic within each epoch with five critic iterations to every generator iteration. This ensures the critic is providing high-quality information to the generator. We initially train the generator to convergence without the ambiguity loss, and then fine tune for a smaller number of epochs with both loss terms. We train the generator using the Adam optimizer with a learning rate of $1 \times 10^{-5}$, and train the critic using root mean square propagation (RMSProp) with a learning rate of $5 \times 10^{-5}$. We use a global batch size of 512 and train on two NVIDIA Tesla V100 GPUs.

\subsection{Ambiguity Loss} 
\label{section::AmbiguityLoss}

The WGAN critic enforces LPD performance by minimizing the distance between the RF background and generated distributions. We then include an \textit{ambiguity loss} to simultaneously optimize the generated waveforms for desirable ambiguity characteristics. We define these desired ambiguity characteristics with a target ambiguity function
\begin{equation}
  \hat{A}_T(\tau, F_D) =
  \begin{cases} 
      \chi & \text{if } (\tau, F_D) = (0, 0) \\
      0 & \text{otherwise}
  \end{cases}
  ,
\end{equation}
which is an ideal ``thumbtack'' ambiguity (Fig. \ref{thumbtack}) with a variable target peak $\chi$.

Our ambiguity loss is composed of two parts. First, we enforce the condition that the generated mainlobe amplitude must be \textit{at least} as large as the target mainlobe:
\begin{equation}
    L_{\text{main}} = \frac{1}{M}\|\text{ReLU}(\hat{A}_T(0, 0) - \hat{A}_G(0, 0))\|_2^2,
\end{equation}
where $M$ is batch size of the generated waveforms and $\hat{A}_G$ is a batch of ambiguity images computed from the generated waveforms $G(z_m|y_m)$. This does not penalize the network for a larger mainlobe amplitude. We then enforce the condition that all sidelobes should be pushed to zero, with increased weight for the zero-Doppler slice:
\begin{equation}
    L_{\text{side}}(\tau, F_D) = \frac{1}{M}\|\hat{A}_T(\tau, F_D) - \Gamma \circ \hat{A}_G(\tau, F_D)\|_2^2,
\end{equation}
evaluated for all $(\tau, F_D) \neq (0,0)$. $\Gamma$, or the \textit{zero-Doppler weight}, is a weight matrix that increases the loss for the zero-Doppler slice by a factor of $\gamma$:
\begin{equation*}
    \Gamma =
    \begin{bmatrix}
    1 & \dots & 1 \\
    \vdots & \ddots & \vdots \\
    1 & \dots & 1 \\
    \gamma & \dots & \gamma \\
    1 & \dots & 1 \\
    \vdots & \ddots & \vdots \\
    1 & \dots & 1 \\
    \end{bmatrix},
\end{equation*}
which prevents optimizing to minima with low Doppler sidelobes and high zero-Doppler sidelobes. The mainlobe loss and sidelobe loss terms are then combined for our full ambiguity loss:
\begin{equation}
    \label{ambig_loss}
    L_{\text{ambig}} = L_{\text{main}} + L_{\text{side}}(\tau, F_D)\big|_{(\tau, F_D) \neq (0,0)}.
\end{equation}
We compute the ambiguity loss for 256 Doppler slices over a $\pm$10\% Doppler shift, yielding ambiguity images with $1024 \times 257$ resolution. Combining the ambiguity loss with the cWGAN-GP loss gives our total objective function
\begin{equation}
    \label{total_loss}
    L_{\text{total}} = L_W + \eta L_{\text{ambig}},
\end{equation}
where $\eta$ is a scaling parameter to set the strength of the ambiguity loss. Larger values of $\eta$ will force the generator to produce waveforms that are less ambiguous.

\section{Experimental Setup}
\label{section::Setup}

We evaluate our generator network by testing its output diversity, detectability, and ambiguity characteristics. We first train the network on a simple, well-defined background distribution to test its diversity. We then train the network on a complex RF background and compare detectability and ambiguity characteristics of the generated waveforms against traditional LPD waveforms.

\subsection{RF Background} 
\label{section::RFBackground}

\begin{table}[t]
    \caption{Overview of SIDLE radar dataset \cite{Rigling2010}}
    \begin{center}
    \begin{tabular}{ c c c c }
    \hline \hline
    Class & Modulation Type & Code Length & Pulse Width ($\mu \text{s}$) \\ 
    \hline
    1 & Barker & 7 & [0.875, 7] \\  
    2 & Barker & 11 & [1.375, 11] \\
    3 & Barker & 13 & [1.625, 13] \\
    4 & Combined Barker & 16 & [1, 8] \\
    5 & Combined Barker & 49 & [3.08, 21.1] \\
    6$^{*}$ & Combined Barker & 169 & [10.58, 84.6] \\
    7$^{\dag}$ & Max. Len. Pseudo Rand. & 15 & [1, 4.5] \\
    8$^{\dag}$ & Max. Len. Pseudo Rand. & 31 & [0.235, 10.5] \\
    9$^{\dag}$ & Max. Len. Pseudo Rand. & 63 & [4.221, 18.9] \\
    10 & Min. Peak Sidelobe & 7 & [1.05, 4.2] \\
    11 & Min. Peak Sidelobe & 25 & [1.25, 10] \\
    12 & Min. Peak Sidelobe & 48 & [2.4, 19.2] \\
    13$^{\dag}$ & T1 & N/A & [2, 16] \\
    14$^{\dag}$ & T2 & N/A & [1.5, 12] \\
    15$^{\dag}$ & T3 & N/A & [1, 8] \\
    16$^{\dag}$ & Polyphase Barker & 7 & [0.875, 7] \\
    17$^{\dag}$ & Polyphase Barker & 20 & [1, 8] \\
    18$^{\dag}$ & Polyphase Barker & 40 & [2, 16] \\
    19$^{\dag}$ & P1 & N/A & [5, 20] \\
    20$^{\dag}$ & P2 & N/A & [3.2, 25.6] \\
    21$^{\dag}$ & P3 & N/A & [3.2, 25.6] \\
    22$^{\dag}$ & P4 & N/A & [5, 20] \\
    23 & Minimum Shift Key & 63 & [2, 18.9] \\
    \hline \hline
    \multicolumn{4}{l}{$^{*}$Class omitted from experiments}
    \text{$^{\dag}$LPD modulation}
    \label{SIDLETable}
    \end{tabular}
    \end{center}
\end{table}

Let $\mathbb{P}_b$ define the RF background distribution. In realistic scenarios, this distribution can be quite complex, encompassing a variety of RF emitters and their environmental interactions. A complex background distribution provides our generator with a high number of degrees of freedom for its generated distribution, potentially leading to more effective LPD performance. However, this complex distribution can be difficult to quantitatively evaluate. Because of this, we test our generator with two representative cases: (1) a simple toy dataset of linear frequency modulated (LFM) waveforms, and (2) a complex dataset of realistic communication waveforms.

We initially define the RF background using a simple toy dataset of simulated LFM waveforms, or linear ``chirps,'' so that we can understand the generated distribution's characteristics in a controlled setting. A linear chirp is defined as
\begin{equation}
s_{\text{chirp}}(t) = A\exp(j \pi \beta t^2 / T ),
\end{equation}
where $A$ is the amplitude, $\beta$ is the sweep bandwidth, and $T$ is the pulse duration. We generate our chirp dataset with examples defined as
\begin{equation}
x_{\text{chirp}}(t) = \exp(j \pi B t^2)+n(t),
\end{equation}
where $B$, the sweep bandwidth, is a discrete random variable drawn from a uniform distribution of integer values $B~\sim~U(5,19)$~Hz. We use a normalized pulse duration $T=1$ second, such that the time-bandwidth product $BT$ of our generated dataset lies in the range of $[5,19]$. We introduce additive white Gaussian noise as $n(t) \sim \mathcal{N}(0,\sigma^2)$. Our generated dataset contains 50,000 waveforms with 1024 IQ time samples at 30 dB SNR. We use this simple chirp dataset to demonstrate the distribution-matching capabilities of our WGAN. 

We then formulate a more complex problem by defining the RF background distribution with common communication waveforms from the DeepSig RadioML 2018.01A dataset \cite{RadioML2018}. This dataset contains over-the-air recordings of the following 24 digital and analog modulations with synthetic simulated channel effects: 
\begin{quote}
OOK, ASK4, ASK8, BPSK, QPSK, PSK8, PSK16, PSK32, APSK16, APSK32, APSK64, APSK128, QAM16, QAM32, QAM64, QAM128, QAM256, AM\_SSB\_WC, AM\_SSB\_SC, AM\_DSB\_WC, AM\_DSB\_SC, FM, GMSK, and OQPS.
\end{quote}
Each class contains 106,496 waveforms with variable SNRs from --20 to 30 dB in increments of 2 dB. Each waveform consists of 1024 IQ time samples, and SNR is calculated based on the average power across all 1024 samples.

\subsection{LPD Effectiveness} 
\label{section::LPDComparison}

\begin{figure}[t]
    \centerline{\includegraphics[width=\columnwidth]{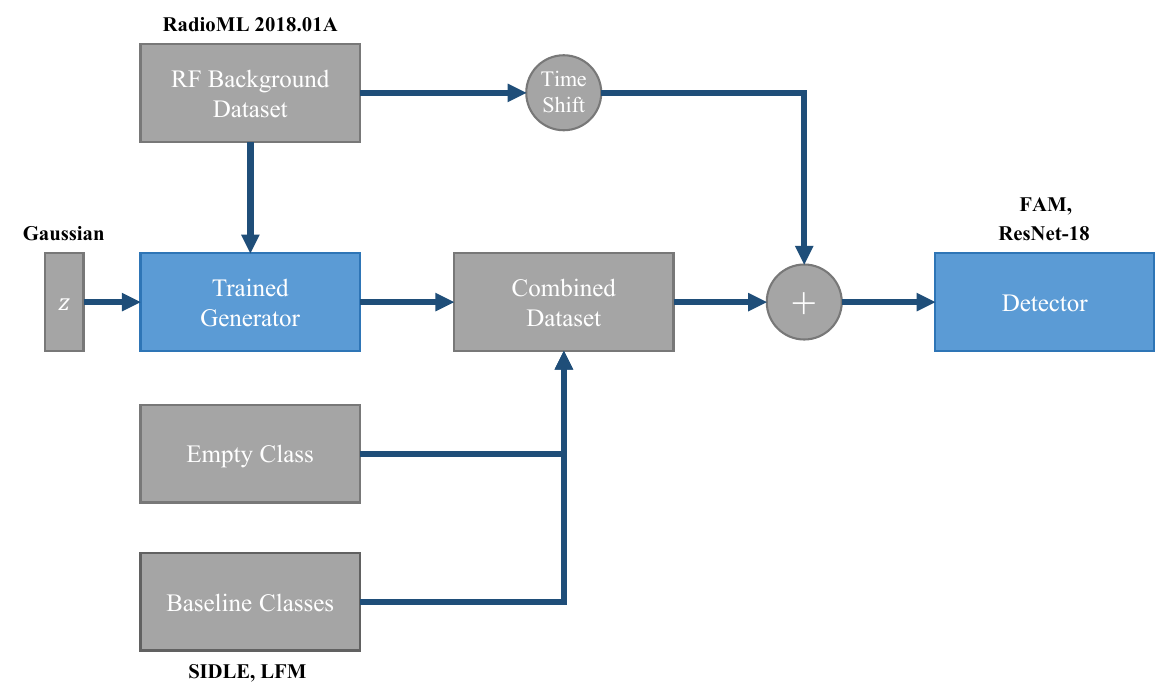}}
    \caption{\textbf{Evaluation Framework.} Training process for the detector neural network used to evaluate LPD performance. We add two additional classes---generated and empty---to the baseline waveforms to create the \textit{combined dataset}. We add randomly time-shifted RF background signals to these examples using waveforms from RadioML 2018.01A that have not been seen by our generator network. We evaluate detectability with an FFT Accumulation Method (FAM)-based cyclostationary detector and a neural network-based detector.}
    \label{LPD-Judge}
\end{figure}

We evaluate the success of our LPD waveform design method by comparing the detectability of our generated waveforms to common radar waveforms from the Air Force Research Laboratory's Signal Instantiations for Deep Learning Experiments (SIDLE) dataset \cite{Rigling2010, Pavy2018, Chakravarthy2020}. This dataset contains simulations of 23 radar modulations, standard and LPD, with variable pulse widths shown in Table \ref{SIDLETable}. The waveforms are sampled at 420 MHz, with time samples per waveform varying from 99 to 10752 samples.
Each class contains 11,000 clean IQ waveforms, and we randomly select a 2.4 $\mu$s (1024 time-sample)
segment from each waveform. For waveforms with less than 1024 samples, we randomly delay the waveform within a 1024 sample frame. We omit class 6 from this dataset as its longer pulse widths required too much memory to use effectively. We add an additional class to this dataset by including a randomized wideband LFM chirp with variable pulse width, offset, and sweep bandwidth ($0.2f_s$--$0.4f_s$).

\begin{figure}
  \centering
  \includegraphics[width=\columnwidth]{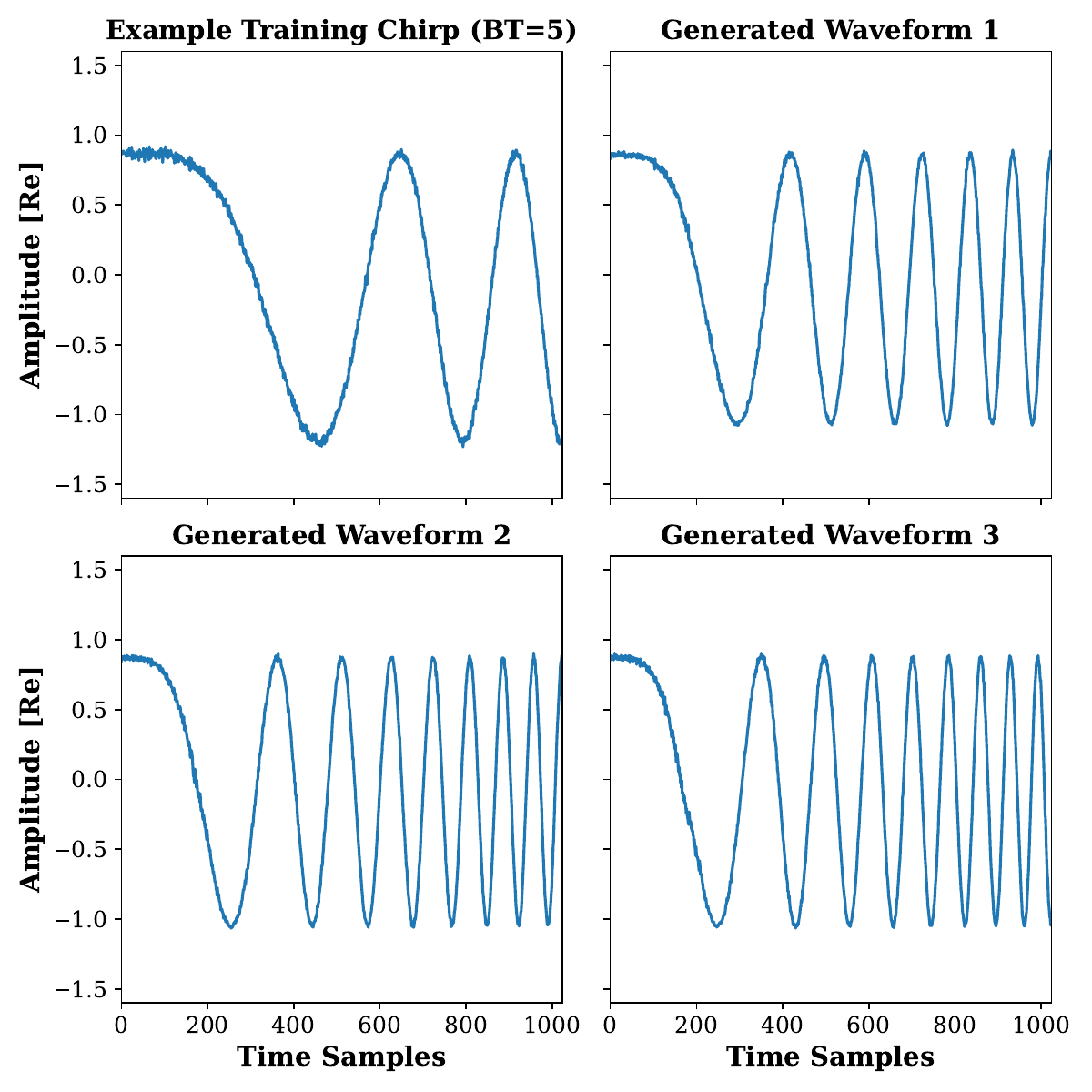}
  \caption{\textbf{Generated Chirp Toy Examples.} We compare a linear chirp from the simulated training dataset (top left) to generated chirps with variable time-bandwidth products from our trained WGAN-GP. Generated examples closely match the distribution of the training dataset.}
  \label{chirp_diversity}
\end{figure}

We compare the detectability of generated and baseline waveforms with two detectors: an FFT Accumulation Method-based cyclostationary detector \cite{Lime2002, FAM2} and a neural network-based detector~\cite{Zhang2017, Ziemann2022, Ma2022}. Both detectors are tasked with detecting the presence of a radar waveform in the presence of the RF background. Radar waveforms are energy-normalized such that they have equivalent detection ranges, and the maximum SNR of the waveforms is 0 dB. Waveforms with pulse widths longer than 2.4 $\mu$s---such as minimum-shift keying and the P-codes---will have lower power with SNRs as low as -10 dB. These longer waveforms are normalized \textit{prior to truncation}, which means the detector will receive less total energy from those waveforms. Power-normalized results with fixed waveform SNR can be found in the Appendix.

Our cyclostationary detector employs the FFT Accumulation Method (FAM) \cite{Lime2002} to detect the cyclostationary properties inherent in radar waveforms. FAM is used to compute an estimate of the spectral correlation function (SCF) which extracts the cyclic features of the signal. Because of the variable cyclostationary characteristics of the RF background, we use an adaptive thresholding detection metric that normalizes the peak SCF value by its time-averaged values. This adaptive threshold triggers a detection when the signal exhibits a higher degree of cyclostationarity than typical, rather than triggering a detection every time a fixed threshold is passed. This performed better than alternative detection metrics (e.g., max- and mean-value) on our data.

Our neural network-based detector uses a ResNet-18 architecture \cite{He2016}. We train the network  on a binary detection problem using a \textit{combined dataset} equally composed of radar waveforms (baseline and generated) and an empty class with no signals present. Examples from the \textit{combined dataset} are energy normalized and combined with a fixed-power, randomly time-shifted RF background signals from the RadioML dataset such that the maximum SNR is 0 dB. Before training the network, inputs are standardized to 0 mean, 1 standard deviation. This process is summarized in Fig. \ref{LPD-Judge}.

We train the ResNet-18 detector with an 80\%/20\% train/test split---sampled equally across SNRs and classes---and we generate new waveforms from our generator with every training and testing epoch. This exposes the detector to the full distribution of generated waveforms $\mathbb{P}_g$, rather than a subset of fixed waveforms. We randomly time shift the RF background signals to ensure our generator does not have oracle time information. We also split the RadioML dataset into two equal parts: one for training our generator and a second for training the detector. This ensures our generator is not trained on any waveforms used to evaluate its LPD effectiveness.

\begin{figure}
  \centering
  \includegraphics[width=.8\columnwidth]{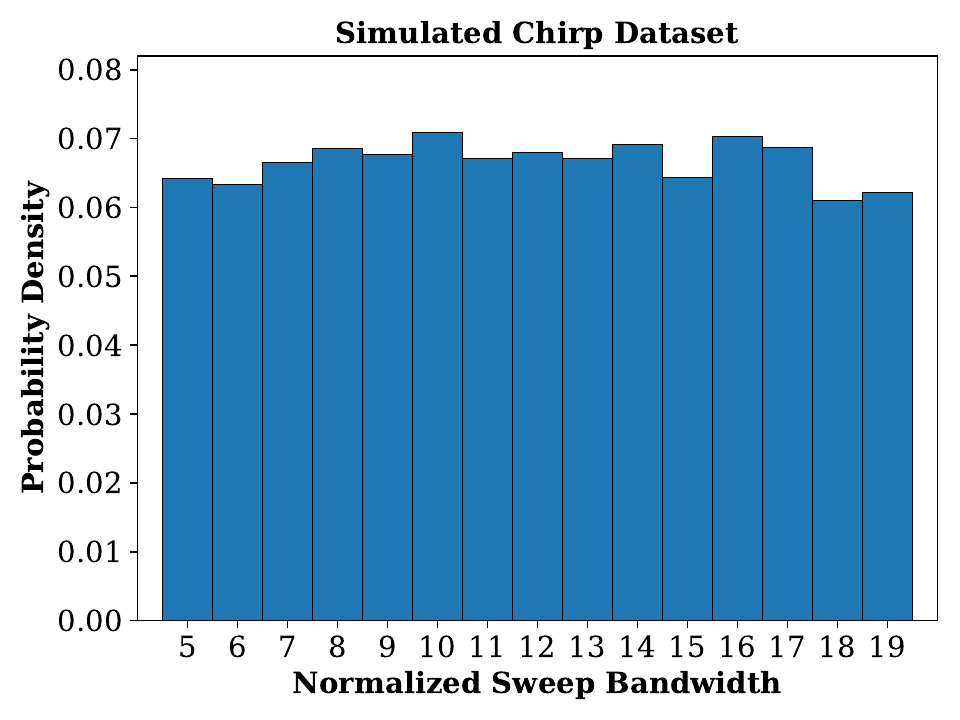}
  \includegraphics[width=.8\columnwidth]{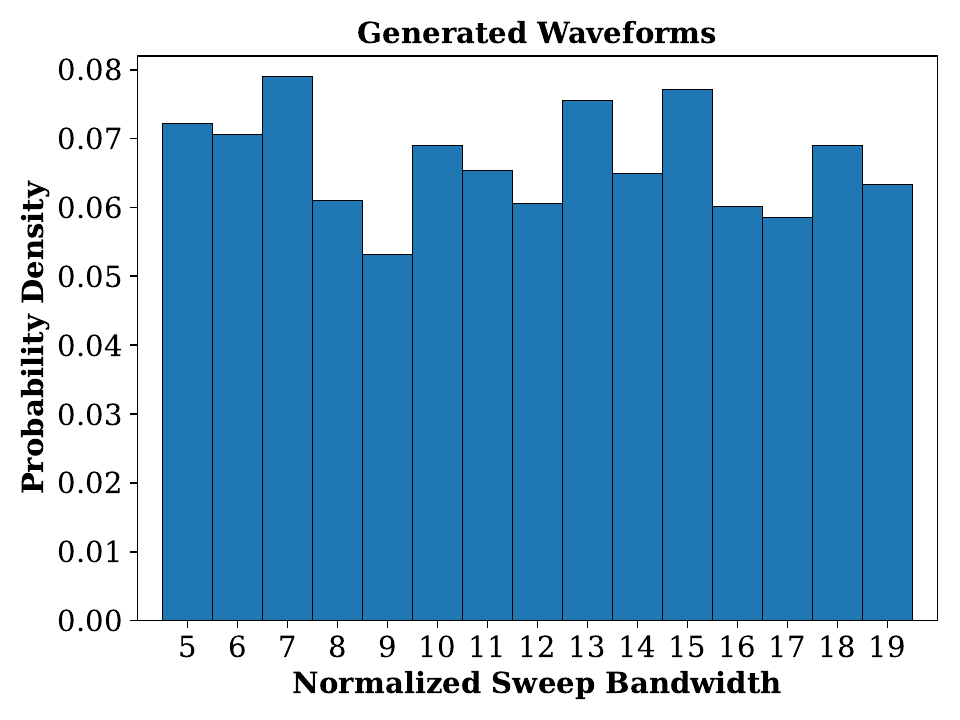}
  \caption{\textbf{Generator Distribution Matching.} Comparison of the sweep bandwidths present within the chirp training dataset (top) and the sweep bandwidths of 5000 waveforms produced by the trained generator (bottom). The training dataset is created with a uniform distribution---the generated distribution closely matches it.}
  \label{chirp_hist}
\end{figure}

\begin{figure}[t]
  \centering
  \includegraphics[width=0.95\columnwidth]{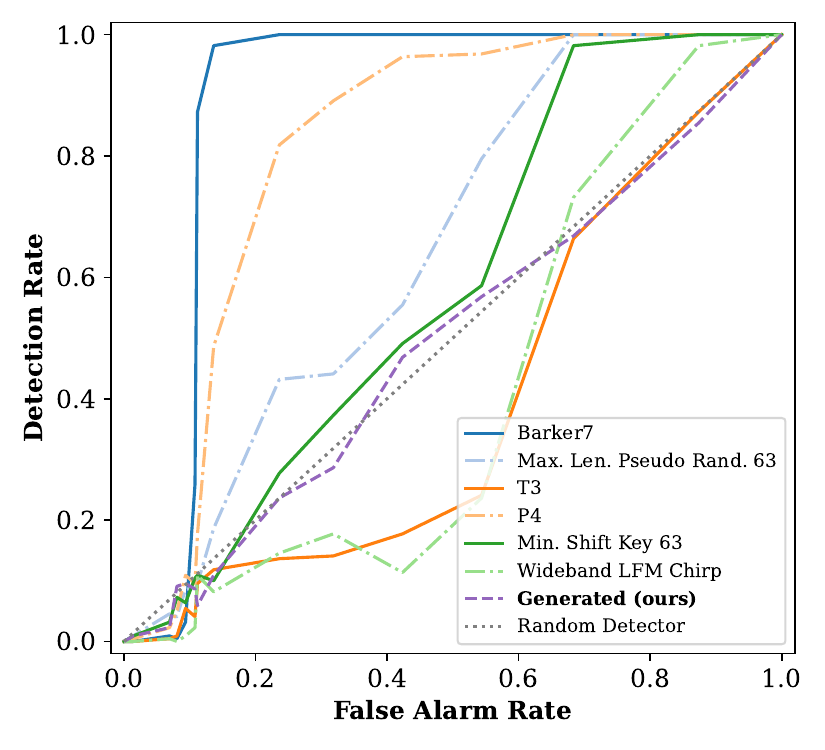}
  \caption{\textbf{FAM ROC Curve.} Receiver operating characteristic (ROC) curves for the FFT Accumulation Method (FAM)-based cyclostationary detector tasked with detecting the baseline waveforms and our generated waveforms in the presence of the RF background. Shown are a subset of baseline waveforms representing the best- and worst-performing. Our generated waveforms closely mirror the random detection chance. This detector fails for low false alarm rates.}
  \label{ROC-FAM}
\end{figure}

\subsection{Sensing Metrics} 
\label{section::Metrics}

We evaluate the sensing performance of our generated waveforms by comparing their ambiguity characteristics to the baseline waveforms. The goal of our ambiguity loss is to optimize our generated waveforms toward the ideal ``thumbtack'' ambiguity. The resulting waveforms should have low ambiguity in delay (range) and Doppler (velocity) with low sidelobes. We use three ambiguity function metrics to evaluate sensing performance: mainlobe width, peak sidelobe level (PSL), and integrated Doppler ambiguity. The mainlobe width and PSL evaluate performance in the zero-Doppler regime, and the integrated Doppler evaluates the overall Doppler ambiguity response of the waveforms.

Mainlobe width ($\Phi_{MW}$) describes the half-power width of the ambiguity mainlobe, and it is an important metric for radar applications \cite{Richards2010}. A narrower mainlobe (central peak) allows for better resolution in range and velocity, thus making it easier to distinguish between closely spaced targets. We define mainlobe width as
\begin{equation}
    \Phi_{MW} \Bigl[\hat{A}(\tau, F_D)\big|_{F_D=0}\Bigr] = \tau_2 - \tau_1,
\end{equation}
where $\tau_1$ and $\tau_2$ are given by
\begin{equation}
    \hat{A}(\tau_1, 0) = \hat{A}(\tau_2, 0) = \frac{1}{2} \cdot \hat{A}(0, 0)
\end{equation}
for $\tau_{1} \in \bigl[-\tau_{\text{main}},0]$ and $\tau_{2} \in \bigl[0,\tau_{\text{main}}]$. Here, the interval $[-\tau_{\text{main}},\tau_{\text{main}}]$ contains the full mainlobe centered on $\tau=0$.

PSL ($\Phi_{PSL}$) is the ratio of the highest amplitude sidelobe to the amplitude of the mainlobe \cite{Blunt2016}. High sidelobes can cause unwanted interference and mask weaker signals near strong reflectors, so a high PSL can lead to decreased detection capability for objects of interest. PSL is defined as
\begin{equation}
    \Phi_{PSL} \Bigl[\hat{A}(\tau, F_D)\big|_{F_D=0}\Bigr] = \max_{\tau}\biggl|\frac{\hat{A}(\tau,0)}{\hat{A}(0,0)}\biggr|
\end{equation}
for $\tau \in \bigl[\tau_{\text{main}},T\bigr]$, where the interval $[\tau_{\text{main}}, T]$ contains all sidelobes. 

\begin{figure}[t]
  \centering
  \includegraphics[width=0.95\columnwidth]{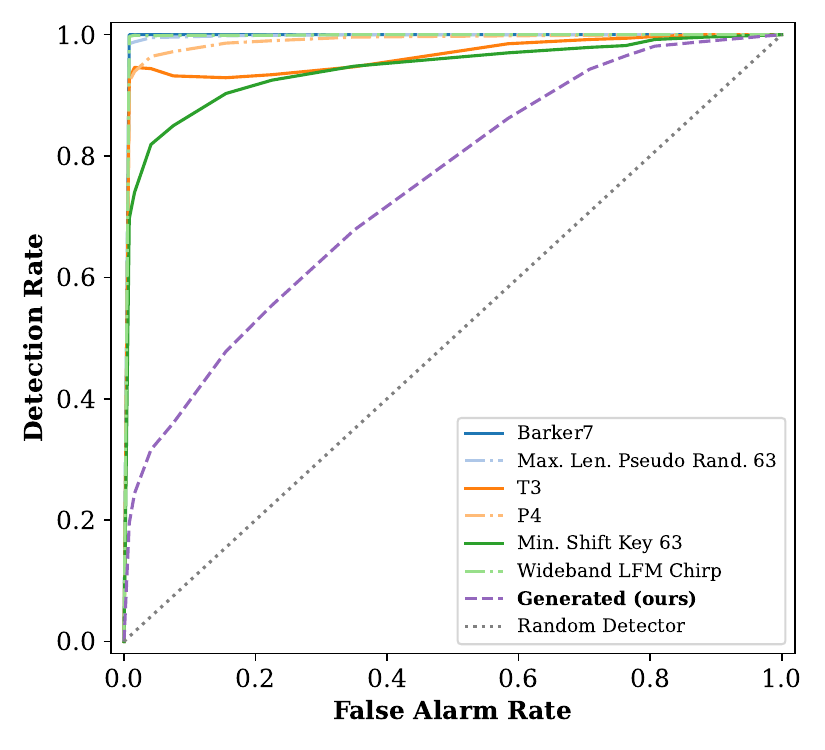}
  \caption{\textbf{ResNet-18 ROC Curve.} Receiver operating characteristic (ROC) curves for the supervised binary detection network tasked with detecting the baseline waveforms and our generated waveforms in the presence of the RF background. Shown are a subset of baseline waveforms representing the best- and worst-performing. Our generated waveforms are strictly harder to detect.}
  \label{ROC-NN}
\end{figure}

Integrated Doppler ambiguity ($\Phi_{ID}$) is the normalized sum of all Doppler terms of the ambiguity function. While not a standard metric, we use it to compare the global Doppler ambiguity response of our waveforms. A larger integrated Doppler value means a waveform has greater Doppler ambiguity. We define integrated Doppler ambiguity as
\begin{equation}
    \Phi_{ID} \Bigl[\hat{A}(\tau, F_D)\Bigr] = \sum_{{\tau = -T}}^{T} \sum_{{\substack{F_D = -F \\ F_D \neq 0}}}^{F} \left| \frac{\hat{A}(\tau, F_D)}{\hat{A}(0,0)} \right|^2
\end{equation}
where the ambiguity function is computed over the delay and Doppler ranges of $\tau \in \bigl[-T,T\bigr]$ and $F_D \in \bigl[-F,F\bigr]$, respectively.

\section{Experimental Results} 
\label{section::Results}

\subsection{Qualitative Example} 
\label{section::Distribution}

We first explore the qualitative behavior of our generative network to ensure it can accurately learn RF distributions. We train the WGAN-GP on our linear chirp dataset with the non-conditional WGAN-GP objective (Eq. \ref{eq:WGAN-GP}). We remove the ambiguity loss term from training (Eq. \ref{ambig_loss}) to isolate performance under the WGAN-GP objective. 

Once trained, the WGAN-GP accurately reproduces examples from the linear chirp distribution. Examples of generated waveforms can be seen in Fig. \ref{chirp_diversity}. To verify the generated distribution, we extract the time-bandwidth product from 10,000 generated chirps and compare them in Fig. \ref{chirp_hist}. The simulated chirp dataset was generated with a uniform distribution of integer time-bandwidth products in the range $[5,19]$, and the generated distribution closely matches that uniform distribution.

\subsection{Detectability} 
\label{section::Detectability}

We then extend our framework to the complex RF background dataset, RadioML 2018.01A, to evaluate LPD performance. We have shown the network can effectively minimize the distance between training and generated RF distributions. Since this distance is an effective surrogate for detectability, our generated waveforms should be more difficult to detect.

We test this hypothesis by training our conditional WGAN-GP on the 50\% training split of RadioML with the cWGAN-GP loss (Eq. \ref{eq:cWGAN-GP}). We omit the ambiguity loss to isolate the effects of the cWGAN-GP loss term. We compare the performance of our FAM-based cyclostationary detector and neural network-based detector on detecting our generated waveforms and the baseline waveforms. These waveforms are energy-normalized and summed to the 50\% test split of RadioML. Power-normalized results at fixed waveform SNR can be found in the Appendix.

We find that our generator is able to produce effective LPD waveforms. We compare the receiver operating characteristic (ROC) curves for both the FAM and ResNet-18 detectors, shown in Fig. \ref{ROC-FAM} and Fig. \ref{ROC-NN} respectively. The ResNet-18 detector is trained on an equal number of examples for baseline and generated waveforms (8800 per class). To increase the false positive rate for the ResNet-18, we retrain the detector multiple times with a decreasing ratio of empty labels in the training dataset.

We find that the cyclostationary detector performs poorly in this detection scenario, which is expected due to the complex cyclostationary RF background. Our generated waveforms are detected at a rate of random chance at high false alarm rates, and the T3 and minimum-shift keying classes are detected at below random chance. However, this detector fails at the low false alarm rates we are most interested in, with all classes degrading to below random chance.

The ResNet-18 detector performs significantly better, especially at low false alarm rates. Nearly all baseline classes are detected at a rate of over $95\%$ at a $1\%$ false alarm rate (FAR), and that rate quickly rises to over $99\%$. The exception is the minimum-shift keying class, which is detected at a rate of $70\%$ at $1\%$ FAR. Notably, our generated waveforms are detected at a rate of $19.5\% \pm 5.5\%$ at $1\%$ FAR, and they remain more difficult to detect than the baseline waveforms at all false alarm rates.

\begin{figure}
  \centering
  \includegraphics[width=0.95\columnwidth]{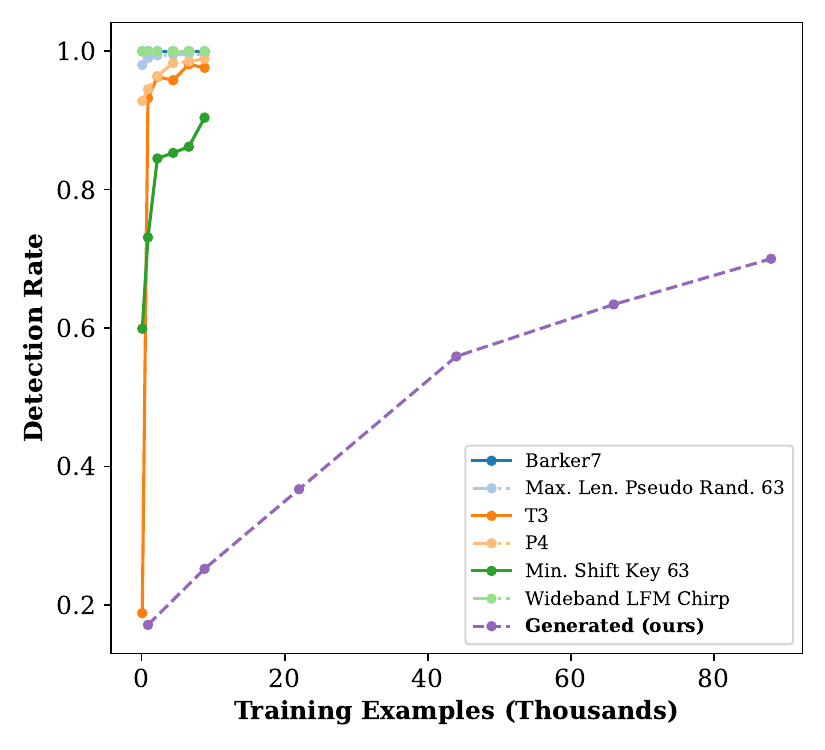}
  \caption{\textbf{Detection Rate vs Data Quantity.} ResNet-18 detection rate comparison for selected baseline waveforms and our generated waveforms with variable training quantity. Generated waveforms significantly outperform the baseline waveforms even with 10$\times$ more training data.}
  \label{acc_data}
\end{figure}

The ResNet-18 detector is supervised, and so its performance is dependent on the amount of training examples seen by the detector, as shown in Fig. \ref {acc_data}. For an equal number of training examples as the baseline classes ($8800$), our generated waveforms maintain an average detection rate of $24.4\% \pm 5.6\%$. Of the baseline classes, the minimum-shift keying class performs the best at a rate of $74\%$---all other waveforms are detected at a rate of over $95\%$, with most over $99\%$. 

When the detector is exposed to 10$\times$ more generated data (88,000 training examples), detection rate is increased to $70.0\%$. This is expected, as increased exposure to generated waveforms will allow the detector to more accurately learn the generated distribution. However, this behavior plateaus, with 100$\times$ more generated data yielding an $80.8\%$ detection rate. This shows that the generated waveforms maintain improved LPD performance even with significantly more training data.

\subsection{Sensing Performance} 
\label{section::Performance}

We fine tune our generator for 3,000 iterations with both the cWGAN-GP loss and ambiguity loss (Eq. \ref{ambig_loss}) to improve the ambiguity characteristics of the generated waveforms. We evaluate the effectiveness of the ambiguity loss term by computing the mainlobe width (MW), peak sidelobe level (PSL), and integrated Doppler ambiguity (IDA) of the generated waveforms conditioned on all RF background classes at $\{-10,0,10,30\}$ dB SNR. We compare these to the same metrics computed for the baseline waveforms. Because of the variable signal length of the SIDLE waveforms, we zero pad all waveforms to 11,000 time samples prior to computing their ambiguity function. 

\begin{figure*}
    \centering
    \includegraphics[height=0.215\textheight]{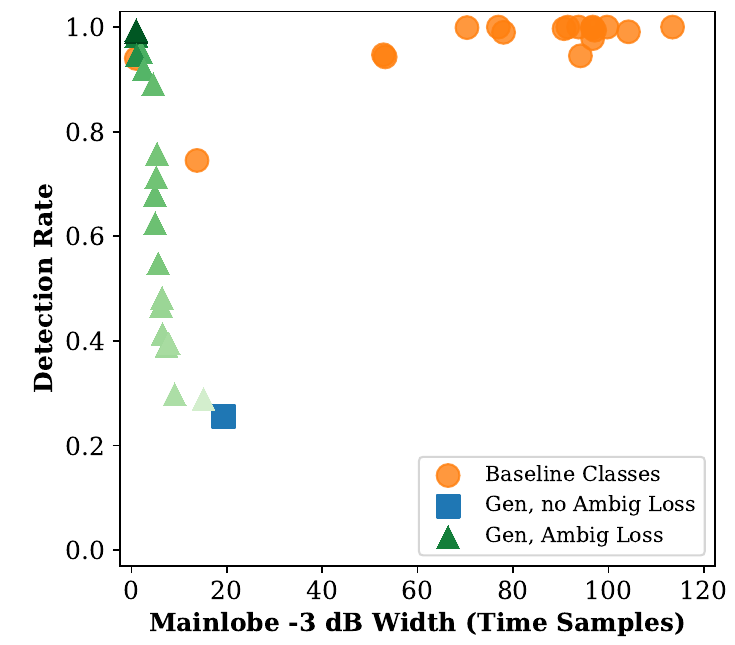}
    \includegraphics[height=0.215\textheight]{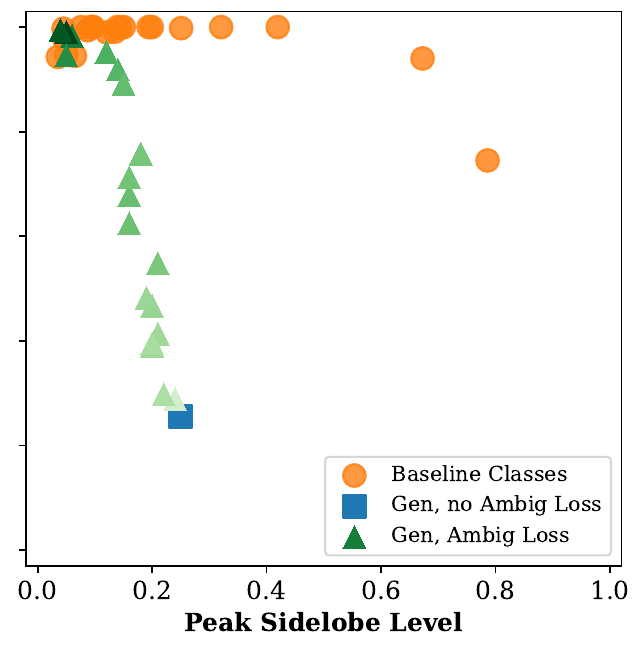}
    \includegraphics[height=0.215\textheight]{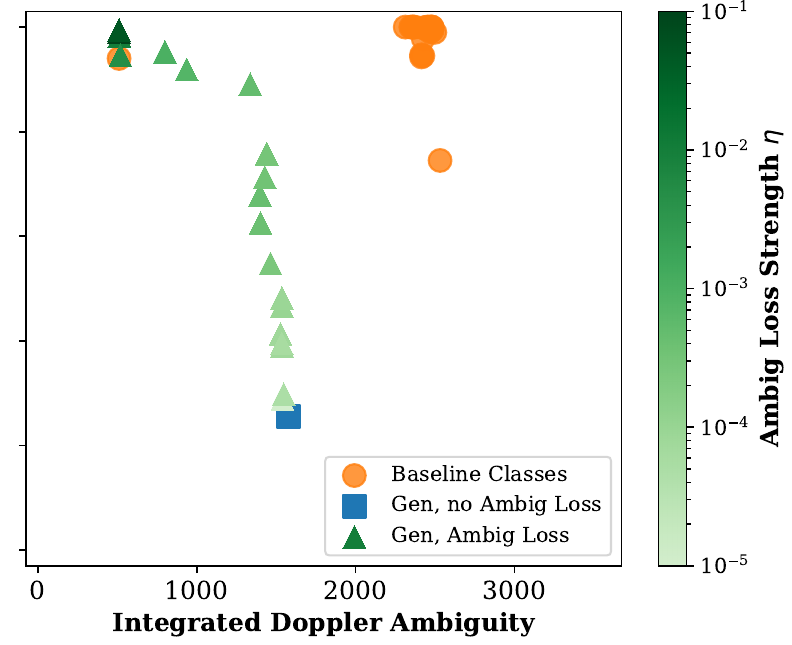}
  \caption{\textbf{Tunable Sensing Performance.} Shown are detectability vs sensing metrics for LPD generators that have been fine-tuned with variable ambiguity loss strengths (shades of green). They are compared to the baseline classes (orange), and a baseline generator without ambiguity loss fine-tuning (blue). Leftmost plot's x-axis represents mainlobe width---lower implies better range resolution. Center plot's x-axis represents peak sidelobe level---lower implies better target separation. Rightmost plot's x-axis represents integrated Doppler ambiguity---lower implies less velocity ambiguity. Adjusting the ambiguity loss strength ($\eta$) allows one to control the trade-off between sensing ability and detectability. Increasing the ambiguity loss strength yields a less ambiguous waveform with narrower mainlobes, improved peak sidelobe level, and lower Doppler ambiguity at the cost of increased detectability. All values of ambiguity loss strength yield waveforms that offer comparable or improved sensing metrics compared to the baseline classes while being harder to detect.}
  \label{fig:sensing_metrics}
\end{figure*}

\begin{figure*}[t]
  \centering
  \includegraphics[height=0.2\textheight]{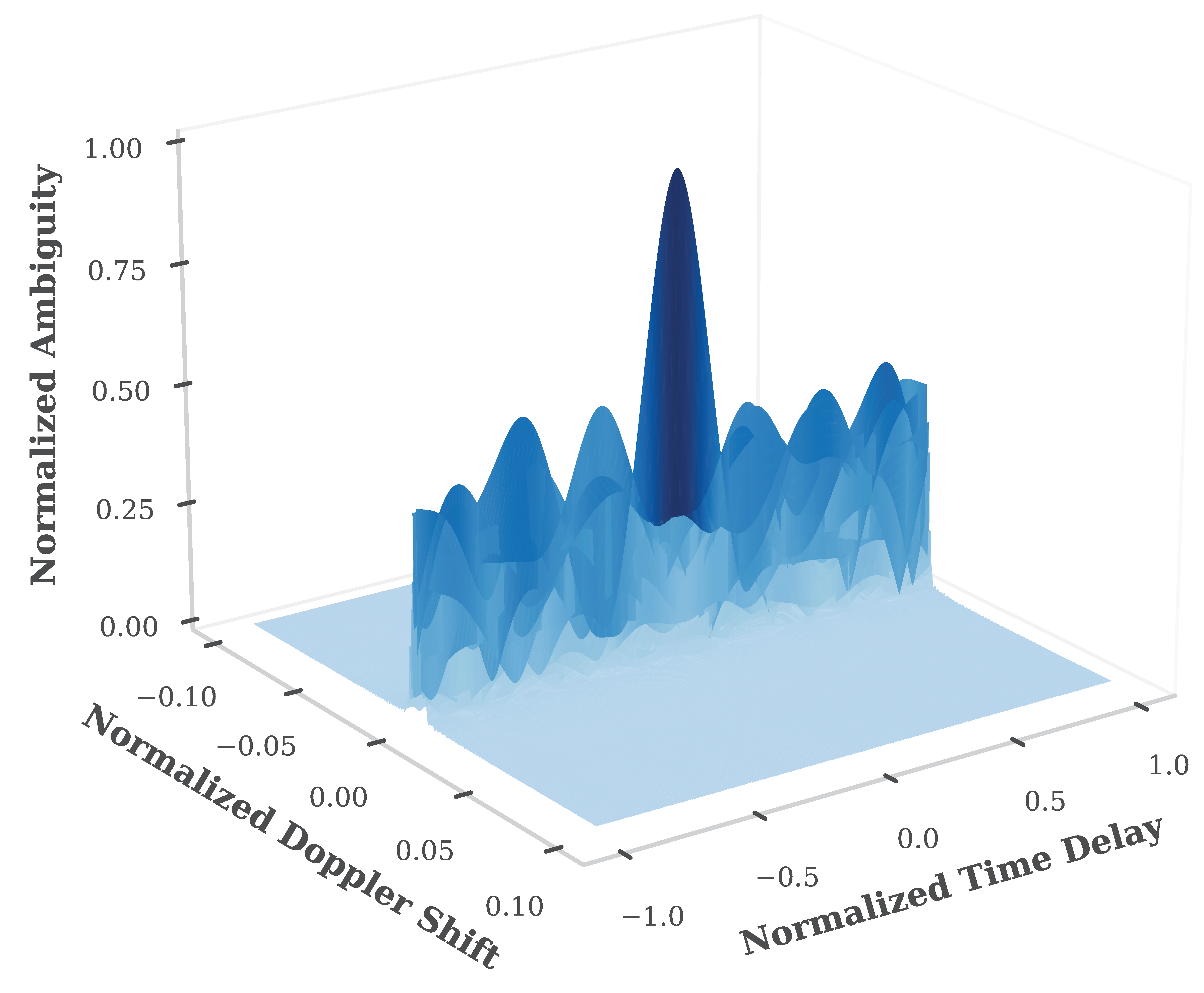}
  \includegraphics[height=0.2\textheight]{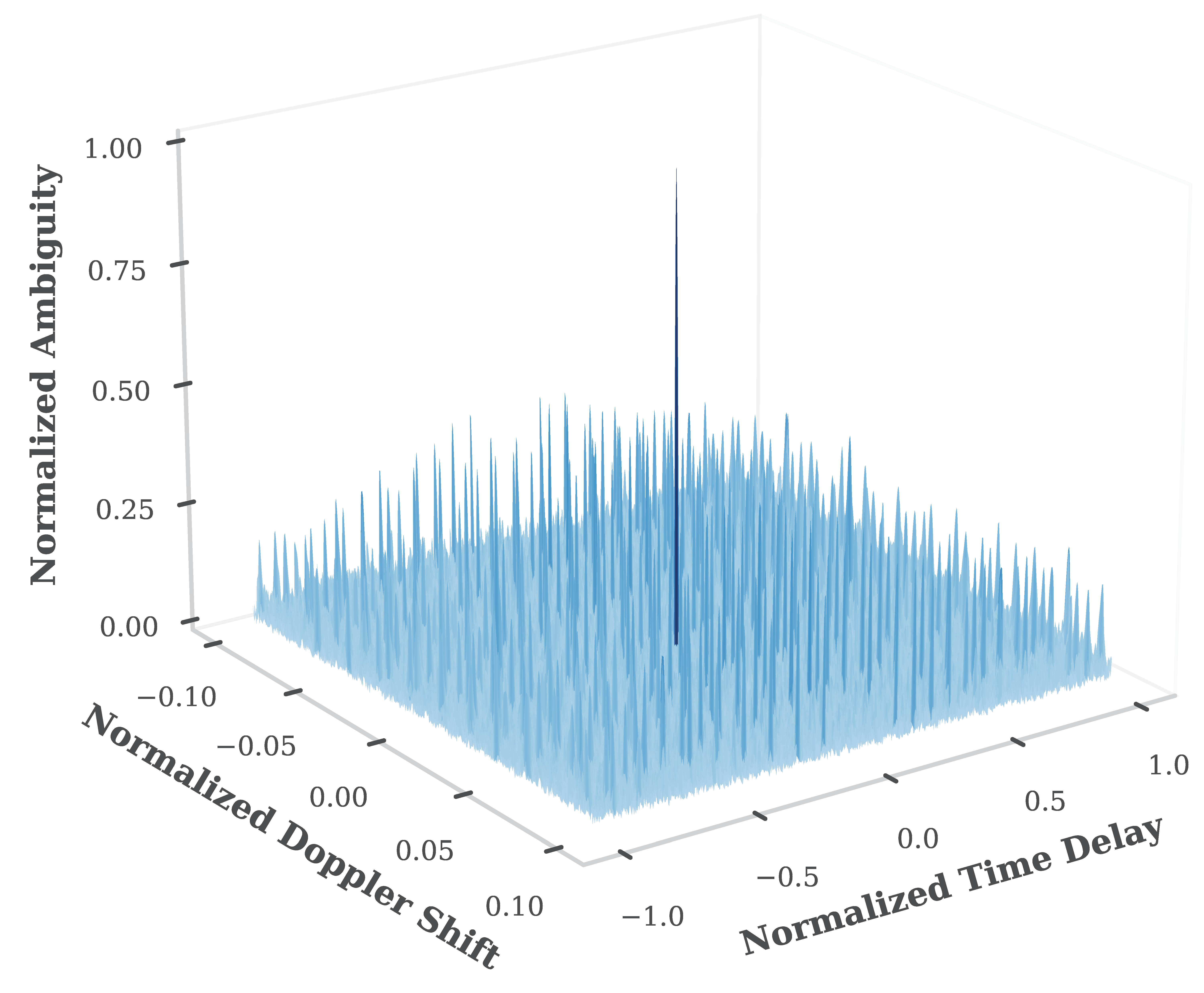} 
  \includegraphics[height=0.2\textheight]{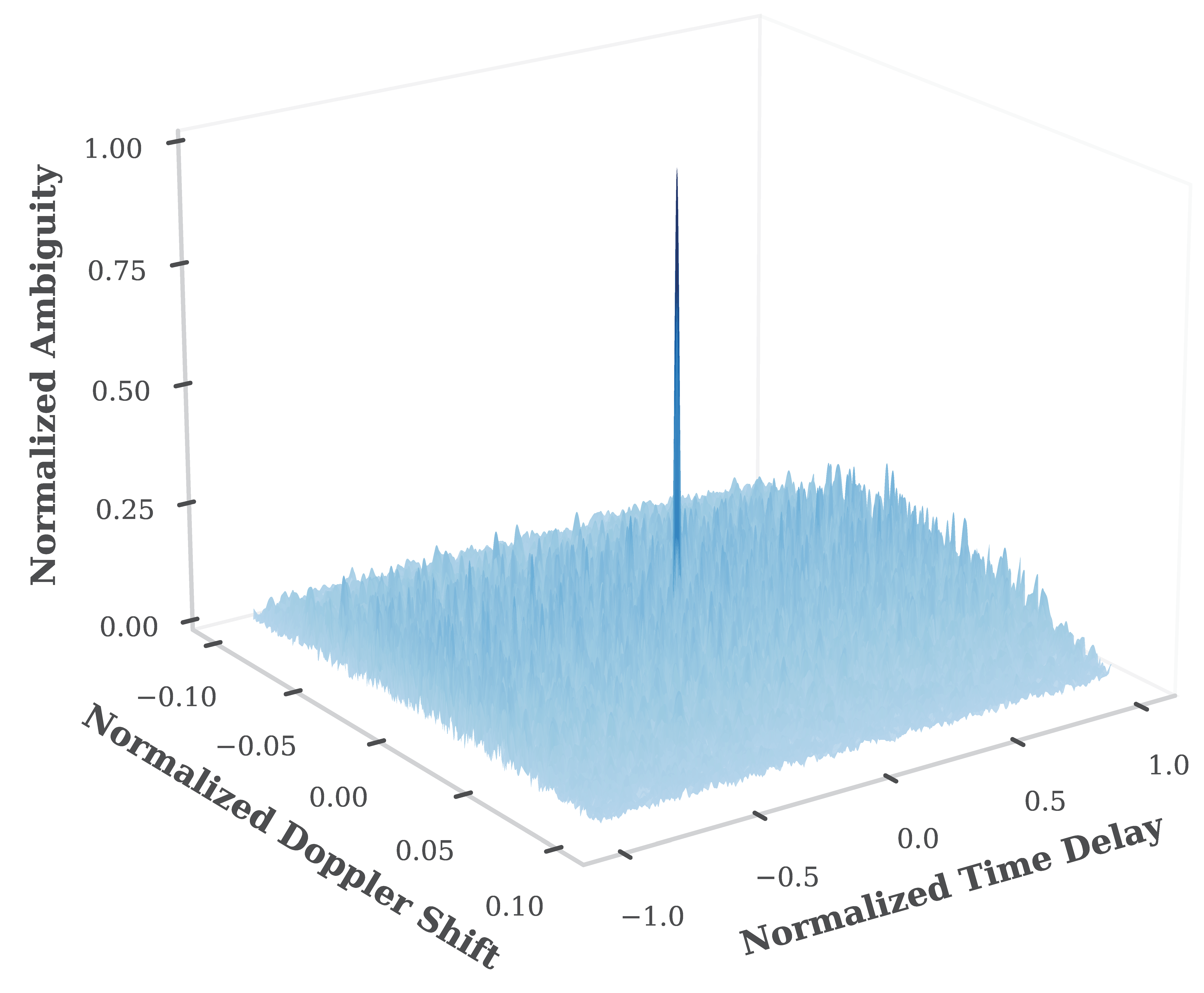}
  \caption{\textbf{Ambiguity Comparison.} Example ambiguity images from SIDLE Barker 7 (left), SIDLE T3 (center), and our LPD waveform generator (right). The generated waveform is created with ambiguity loss strength $\eta=0.0001$ conditioned on a BPSK background signal at 10 dB SNR. The generated waveform retains a narrow mainlobe with low sidelobes for improved range/velocity resolution.}
  \label{ambig_comparison}
\end{figure*}

We find that the ambiguity loss improves the MW, PSL, and IDA of the generated waveforms. To understand this in the context of LPD, we evaluate each model's detectability with the ResNet-18 detector trained on an equal number of baseline and generated examples (8800 per class). The resulting detectability is shown as a function of MW, PSL, and IDA in Fig.~\ref{fig:sensing_metrics}. The generators used in these figures were fine-tuned with a variable ambiguity loss strength $\eta$, target peak $\chi=1000$, and zero-Doppler weight $\gamma=1000$. Larger $\eta$ leads to improved MW, PSL, and IDA at the cost of greater detectability. However, generated waveforms with no ambiguity loss retain MW, PSL, and IDA values comparable or better than those of the baseline waveforms, and match the values associated with the RF background. A comparison of ambiguity examples from baseline and the LPD generator can be seen in Fig.~\ref{ambig_comparison}.

\section{Discussion}
\vspace{2pt}
\noindent \textbf{Bandwidth Requirements.} Our generated waveforms mimic the ambient background, which in our work consists of communications signals and noise. Accordingly, for low SNR backgrounds our generated signals have a wide bandwidth (similar to the background noise) while for high-SNR backgrounds our generated signals have a relatively narrow bandwidth (similar to the communications signals). For low SNR backgrounds (-20 to 6 dB), our generated waveforms had an average 90\% power bandwidth (the frequency range which contains 90\% of the signal power) of $0.43f_s$, where $f_s$ denotes the sampling rate. Meanwhile for high SNR backgrounds (6 to 30 dB), our generated waveforms had an average bandwidth of $0.06f_s$. For comparison, the SIDLE waveforms had an average bandwidth of $0.04f_s$. In principle, the end-to-end nature of our generation framework allows one to add a third loss term to penalize and constrain the generated waveforms' bandwidth at low SNRs, but we leave this as future work.

\vspace{2pt}
\noindent \textbf{Comparison to White Noise.} A natural comparison to our method is the use of Gaussian noise-based waveforms as in noise radar. Gaussian noise waveforms are a special case of our method when the RF background is dominated by negative SNR signals. In this case, our framework converges to white noise as the optimal LPD waveform. When we compare the detectability of generated waveforms to white noise for \mbox{--20--0} dB SNR RF background signals, generated waveform detectability is 13\% while white noise is below the false positive rate, effectively 0\%. In contrast to noise radar, our generator generalizes to cases where the RF background contains positive SNR signals. For 6--30 dB SNR RF background signals, our generated waveforms achieve a 38.6\% detection rate while white noise is detected at a rate of 92.7\%. 

\vspace{2pt}
\noindent \textbf{Conditional Class Performance.} The distribution-matching nature of our framework necessitates that its outputs are limited by the signals present in the RF background. While we have shown that this is effective for a large number of background signals, not all signals are equally effective. Specifically, we have found that generated LPD waveforms conditioned on AM and FM modulations have lower sensing performance than other classes. Omitting AM and FM conditional classes from our generator yields an overall 35\% improvement in mainlobe width, 12\% improvement in PSL, and  no change in IDA.

\vspace{2pt}  
\noindent \textbf{Background Interference.} Does mimicking the RF environment make our waveform more susceptible to interference at the receiver? To better understand the effects of interference on our waveforms, we re-run our ambiguity analysis with the RF background signal present in the return signal. We interfere our generated and baseline waveforms with the RF background at 0 dB SNR. We find our method degrades similarly to the baseline waveforms under interference. The interfered ambiguity metrics for our generated waveforms consist of an average MW of 23.1 samples (from 19.7), PSL of 0.38 (from 0.25), and IDA of 10984 (from 1576)---lower is better for all metrics. The interfered ambiguity metrics for the baseline waveforms follow a similar trend, with an average MW of 108.2 samples (from 111.1), PSL of 0.3 (from 0.19), and IDA of 42553 (from 2345). Overall, MW is not significantly affected, but PSL is increased and IDA is significantly increased. Notably, the vast majority of degradation in performance for our generated waveforms occurs when generated signals are conditioned on AM and FM modulations. As noted in the previous paragraph, waveforms conditioned on these modulations generally have inferior ambiguity characteristics when compared to other generated waveforms.

\section{Conclusion}
We have developed an unsupervised technique for generating radar waveforms with a low probability of detection. We utilize a generative adversarial framework to design waveforms that mimic the ambient RF environment, making them difficult to distinguish from the RF background. To maintain the sensing capabilities of these waveforms, we introduce and optimize an ambiguity function-based loss metric. Our evaluation demonstrates that the generated waveforms achieve up to a 90\% reduction in detectability when compared to traditional LPD waveforms and retain desirable ambiguity characteristics. Furthermore, we are able to tune the generated waveforms for a desired trade-off between detectability and sensing accuracy.

This generative framework offers substantial versatility in waveform design beyond just LPD. By adjusting both the WGAN's target distribution and the target shape of the ambiguity loss, we can tailor the design to meet specific operational requirements. The inclusion of other objectives---e.g., a bandwidth constraint---may enable further applications such as anti-interference. Given the dynamic landscape of technological advancements and evolving RF conditions, these generative techniques offer a new tool to ensure radar systems remain both effective and adaptive across a wide array of requirements and scenarios.

\appendix
Our presented detection results for both FAM and ResNet-18 detectors utilized energy-normalized data, as this takes radar pulse width into account and ensures all waveforms have the same effective detection range. However, this leads to variable SNRs for all waveforms. Of equal interest from the perspective of detection are power-normalized results, which provides detection behavior at fixed SNRs.

\begin{figure}
  \centering
  \includegraphics[width=0.95\columnwidth]{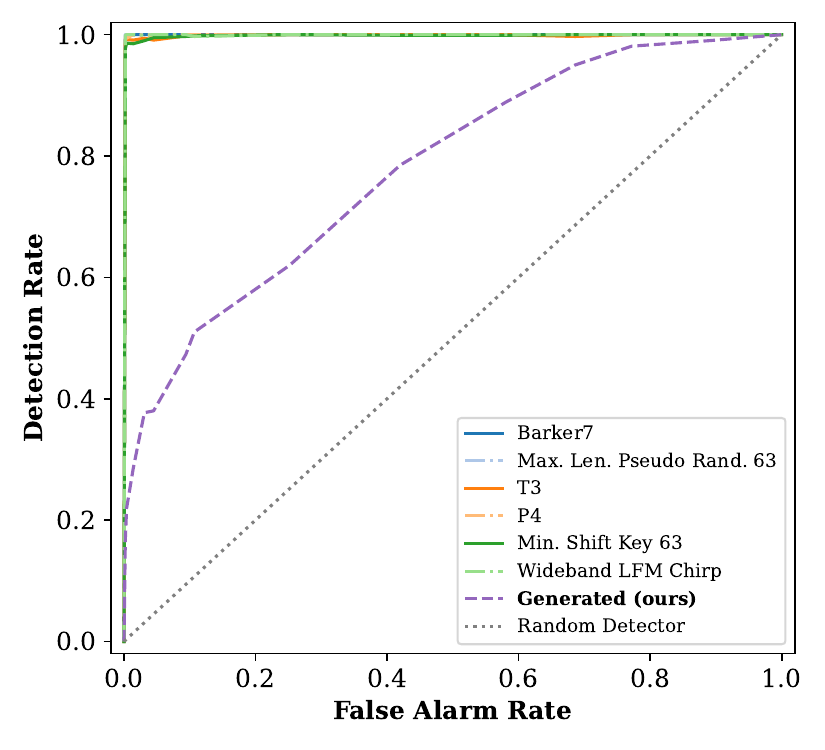}
  \caption{\textbf{Power-normalized ResNet-18 ROC Curve.} Receiver operating characteristic (ROC) curves for the supervised binary detection network on \textit{power-normalized} data. Baseline and generated waveforms are summed with the RF background at a fixed SNR of 0 dB. Our generated waveforms remain strictly harder to detect.}
  \label{ROC-NN-POWER}
\end{figure}

\begin{figure}
  \centering
  \includegraphics[width=0.95\columnwidth]{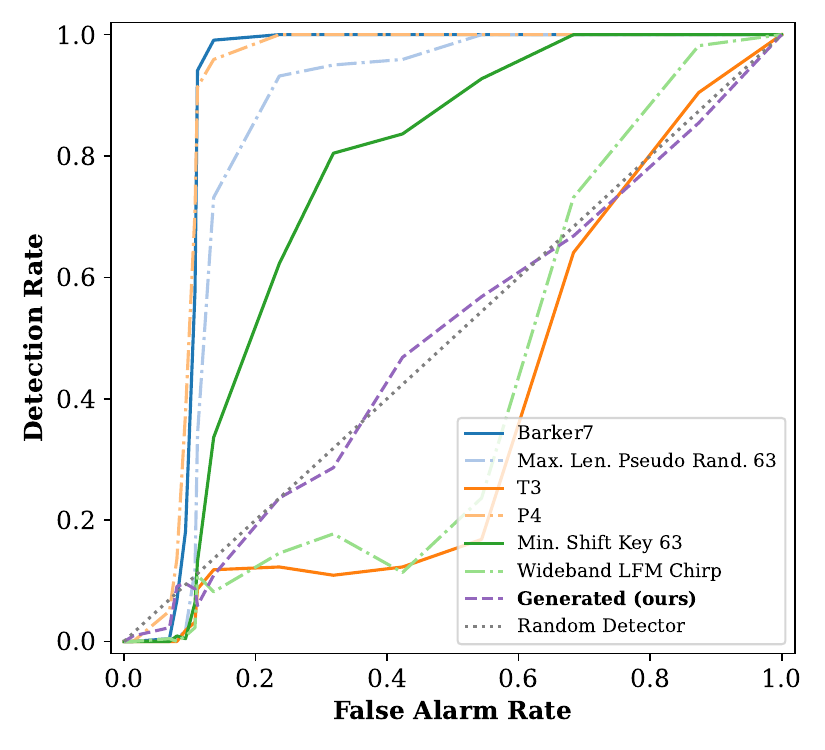}
  \caption{\textbf{Power-normalized FAM ROC Curve.} Receiver operating characteristic (ROC) curves for the FFT Accumulation Method (FAM)-based cyclostationary detector on \textit{power-normalized} data. Baseline and generated waveforms are summed with the RF background at a fixed SNR of 0 dB. Our generated waveforms closely mirror the random detection chance. This detector still fails for low false alarm rates.}
  \label{ROC-FAM-POWER}
\end{figure}

We provide power-normalized detection results for both the ResNet-18 and FAM detectors in Fig. \ref{ROC-NN-POWER} and Fig. \ref{ROC-FAM-POWER} respectively. Signals of interest are fixed to 0 dB SNR. In general, the FAM detector continues to perform poorly at low false alarm rates, though some classes are detected more easily. Notably, the ResNet-18 detector improves significantly at this fixed SNR, detecting all baseline classes at all false alarm rates at over $95\%$. Our generated waveforms continue to perform well in this power-normalized case, retaining a $22\% \pm 5.6\%$ detection rate.

In comparison to the energy-normalized results, these power-normalized results emphasize the importance of temporally spreading waveform energy. Our current framework fixes the pulse width of the generated waveform, but allowing the generator to increase that pulse width could yield further improved LPD characteristics.

\section*{Acknowledgment}
This work was supported in part by high-performance computer time and resources from the DoD High Performance Computing Modernization Program. This work was supported in part by the AFOSR Young Investigator Program Award \#FA9550-22-1-0208 and ONR award \#N000142312752.

The U.S. Government is authorized to reproduce and distribute reprints for governmental purposes notwithstanding any copyright notation thereon. The views and conclusions contained herein are those of the authors and should not be interpreted as necessarily representing the official policies or endorsements, either expressed or implied, of the Dept. of Army or the Office of the Under Secretary of Defense for Research and Engineering (OUSD(R\&E)) or the U.S. Government.

\bibliographystyle{./IEEEtran}
\bibliography{./IEEEabrv,./main.bib}

\newpage
\begin{IEEEbiographynophoto}{Matthew R. Ziemann}
is a physicist at the DEVCOM Army Research Laboratory and a PhD student at the University of Maryland, College Park in the UMD Intelligent Sensing Lab. He received his BS in physics at the University of Illinois in 2016, and received his MS in computer science from the University of Maryland in 2023. 
\end{IEEEbiographynophoto}
\begin{IEEEbiographynophoto}{Christopher A. Metzler}
is an Assistant Professor of Computer Science at the University of Maryland, College Park, where he directs the UMD Intelligent Sensing Lab. He received his BS, MS, and PhD in electrical and computer engineering from Rice University in 2013, 2014, and 2019, respectively. He was a postdoc in the Stanford Computational Imaging Lab 2019--2020. His research develops new systems and algorithms for solving problems in computational imaging and sensing, machine learning, and wireless communications. His work has received multiple best paper awards; he recently received NSF CAREER, AFOSR Young Investigator Program, and ARO Early Career Program awards; and he was an Intelligence Community Postdoctoral Research Fellow, an NSF Graduate Research Fellow, a DoD NDSEG Fellow, and a NASA Texas Space Grant Consortium Fellow.
\end{IEEEbiographynophoto}

\vfill
\end{document}